\documentclass[aps,prapplied,amsmath,amssymb,twocolumn,superscriptaddress, floatfix]{revtex4-2}
\usepackage{graphicx}
\usepackage{booktabs}
\usepackage[english]{babel}
\usepackage{soul}
\usepackage{ragged2e} 
\makeatletter
\renewcommand{\fnum@figure}{Fig.~\thefigure}
\renewcommand{\fnum@table}{Tab.~\thetable}
\makeatother
\usepackage[letterpaper,top=2cm,bottom=2cm,left=3cm,right=3cm,marginparwidth=1.75cm]{geometry}
\usepackage{amsmath}
\usepackage{tikz}
\definecolor{pulsecolor}{HTML}{179C7D}
\usepackage{qcircuit}
\usepackage{braket}
\makeatletter
\newcommand*{\balancecolsandclearpage}{%
  \close@column@grid
  \clearpage
  \twocolumngrid
}
\makeatother
\usepackage[colorlinks=true, allcolors=blue]{hyperref}
\usepackage{cleveref}

\AtBeginDocument{%
}
\usepackage{multirow}

\begin{document}

\title{Stability studies on subtractively-fabricated CMOS-compatible superconducting transmon qubits}
\author{Chawki Dhieb}
\email{chawki.dhieb@emft.fraunhofer.de}
\author{Johannes Weber}
\author{Samuel Taubenberger}
\author{Carla Morán Guizán}
\author{Simon J. K. Lang}
\affiliation{Fraunhofer Institut für Elektronische Mikrosysteme und Festkörpertechnologien EMFT, Munich, Germany}
\author{Zhen Luo}
\affiliation{\mbox{School of Computation, Information and Technology, Technical University of Munich, Munich, Germany}}
\author{Emir Music}
\author{Alwin Maiwald}
\author{Wilfried Lerch}
\author{Lars Nebrich}
\affiliation{Fraunhofer Institut für Elektronische Mikrosysteme und Festkörpertechnologien EMFT, Munich, Germany}
\author{Marc Tornow}
\affiliation{Fraunhofer Institut für Elektronische Mikrosysteme und Festkörpertechnologien EMFT, Munich, Germany}
\affiliation{\mbox{School of Computation, Information and Technology, Technical University of Munich, Munich, Germany}}
\author{Thomas Mayer}
\author{Daniela Zahn}
\author{Rui N. Pereira}
\affiliation{Fraunhofer Institut für Elektronische Mikrosysteme und Festkörpertechnologien EMFT, Munich, Germany}

\author{Christoph Kutter}
\affiliation{Fraunhofer Institut für Elektronische Mikrosysteme und Festkörpertechnologien EMFT, Munich, Germany}
\affiliation{\mbox{Center Integrated Sensor Systems (SENS), Universität der Bundeswehr München, Munich, Germany}}
\date{\today}
\begin{abstract}
Developing fault-tolerant quantum processors with error correction demands large arrays of physical qubits whose key performance metrics (coherence times, control fidelities) must remain within specifications over both short and long timescales. CMOS-compatible subtractive fabrication processes are promising candidates for industrially scalable and mass-produced qubit manufacturing. Here, we investigate the temporal stability of 
superconducting transmon qubits fabricated accordingly. During a single cooldown and over a period of 95 hours, we monitored several parameters for 8 qubits, including coherence times $T_1$ and $T_2^*$, which exhibit fluctuations originating primarily from the interaction between two-level system (TLS) defects and the host qubit. Qubit frequencies remain generally stable with variations confined to a 100\,kHz-interval. By comparison with literature data on lift-off-fabricated qubits, we find that our devices exhibit the same scaling of $T_1$ variance to mean $T_1$,  demonstrating equivalent stability and indicating that TLS-driven $T_1$ fluctuations are insensitive to the choice of fabrication method. To assess long-term stability over multiple cooldowns, we tracked two representative qubits over 10 cooldown cycles spanning more than one year. Within the thermal cycles considered, we observed a cumulative downward shift in qubit transition frequencies of approximately 61\:MHz on average across both qubits, likely caused by progressive aging of the Josephson junction's oxide barrier. In contrast, readout resonator frequencies decreased only marginally. Meanwhile, $T_1$ exhibits cycle-to-cycle fluctuations attributable to TLS spectrum reconfiguration, but maintains a stable baseline value throughout the observation period. Together, these findings validate subtractively fabricated CMOS-compatible qubits as a viable platform for scalable quantum processors, exhibiting competitive stability compared with lift-off devices.
\end{abstract}
\maketitle
\section{Introduction}
Developing a fault-tolerant quantum computer implementing quantum error correction (QEC) requires a large number of physical qubits with consistently high-quality metrics \cite{acharya2024quantum,gambetta2017building}.
To scale to such large quantum processing units (QPUs), it is advantageous to move away from lift-off techniques to facilitate processing of larger wafers, e.g., with diameters of 8 inches and above. Industrial-grade, CMOS-compatible qubit fabrication processes are suitable for large wafers and promise high precision and reproducibility.
Previous works have demonstrated the feasibility of producing superconducting qubit wafers with high yield using only CMOS techniques \cite{mayer20253d,mayer2025cmos,lang2025advancing,van2024advanced}. For QEC, the qubit parameters must also satisfy several requirements: high coherence times to provide enough headroom for QEC cycles, and high readout and gate fidelities allowing precise control \cite{acharya2024quantum,putterman2025hardware, fowler2012surface}. An equally critical performance metric that must be considered is the temporal stability of these parameters. If any of the parameters of a qubit or several qubits fall below the required thresholds during operation or between cooldown cycles, the QPU's performance can be compromised \cite{sheldon2016characterizing, chen2023transmon}. Qubits made with lift-off techniques have been shown to exhibit fluctuations in key parameters, namely $T_1$ and $T_2$ \cite{WMI_TUM_17qubit_2024, burnett2019decoherence, kono2024mechanically, muller2015interacting, tuokkola2025methods}. However, studies on the temporal stability of CMOS-compatible qubits, particularly those fabricated subtractively, are scarce. Comparisons of lifetime stability between different fabrication methods are still missing, and the stability of certain parameters, such as readout metrics, has not been addressed in prior works.
\par In this work, we close this gap by systematically characterizing the temporal stability of subtractively fabricated qubits, using a design with minimal noise sources, on two complementary timescales: within a single cooldown over a period of approximately 95 hours, and across 10 cooldown cycles spanning more than a year. We also assess whether our devices achieve competitive performance and stability compared to conventional lift-off devices and provide, to the best of our knowledge, the first systematic benchmarking of $T_1$ stability between different fabrication methods.


\par Our results suggest that adopting CMOS-compatible subtractive fabrication does not compromise the temporal stability of key qubit parameters compared to lift-off devices, supporting the adoption of industrial fabrication processes for scalable quantum processors. The remainder of the paper is organized as follows: Section II describes the experimental methods, Section III.A presents the results of single-cooldown stability measurements, Section II.B addresses long-term stability across cooldown cycles, and Section IV summarizes the findings and their implications.
\section{Experiment}
\par
\begin{figure}
    \centering
    \includegraphics[width=\linewidth]{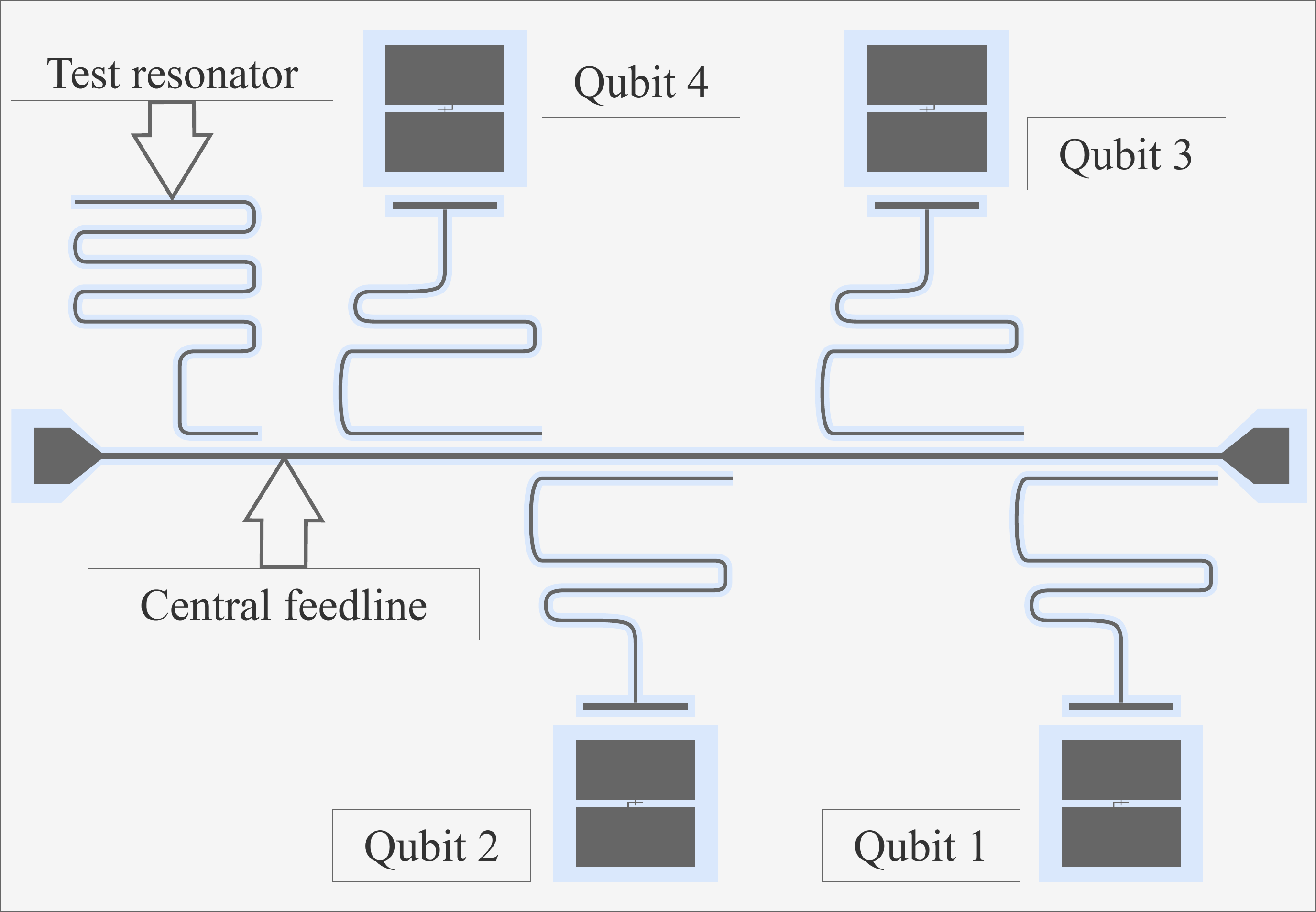}
    \caption{Sketch of a single-qubits chip layout with 4 qubits structures and a test resonator coupled to a central feedline.}
    \label{fig:layout}
\end{figure}

\begin{figure}
    \centering
    \includegraphics[width=0.75\linewidth]{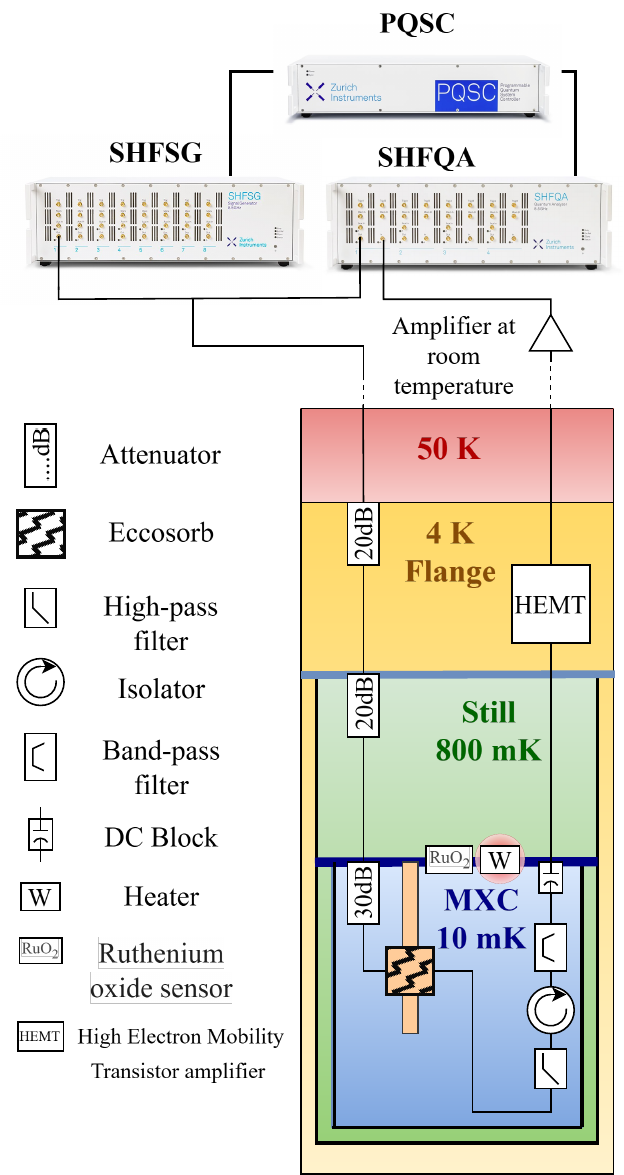}
    \caption{Illustration of the cryogenic measurement setup. The qubit chips are mounted in the mixing chamber (MXC) of a Bluefors LD dilution refrigerator (DR), enclosed in a copper package and covered with Eccosorb. The MXC temperature is monitored with a $\text{RuO}_{2}$-temperature sensor. Drive and input readout signals are generated respectively by a signal generator (SHFSG) and a quantum analyzer (SHFQA), synchronized by a PQSC (all from Zurich Instruments). Both signals go through a combiner at room temperature and then are attenuated at each stage of the DR to suppress thermal noise. The output readout signal passes through filters and isolators before being amplified at the 4K stage and at room temperature.}
    \label{fig:dr}
\end{figure}

 The devices studied here are test transmon qubits that stem from two chips (A and B) from the same wafer, which was fabricated using the method described in \cite{lang2025advancing}. For our test chip layout, we omit dedicated drive lines and rely solely on readout resonators to both drive and read out the qubits. As conceptually depicted in \autoref{fig:layout}, each chip contains 4 qubits plus one reference resonator. \autoref{tab:param_table} summarizes the design parameters of these qubits: readout resonator frequency $f_\mathrm{r}$, qubit frequency $f_\mathrm{q}$, and anharmonicity. The cryogenic setup is described in detail in \autoref{fig:dr}.

\begin{table}
    \centering
    \caption{Frequency parameters of measured qubits}
    \begin{tabular}{c c c c c}
        \toprule
        Chip&Qubit& $f_\mathrm{r}$\,(GHz) & $f_\mathrm{q}$\,(GHz) &Anharmonicity\,(MHz)\\
         \midrule
         A&1&  6.5829& 4.332736& 225.075\\
         A&2&  6.7383& 4.320850& 219.487\\
         A&3&  6.9860& 4.563595& 228.126\\
         A&4&  7.1407& 4.671054& 220.046\\
         \midrule
         B&1&  6.5849& 4.441636& 224.806\\
         B&2&  6.7367& 4.541780& 223.012\\
         B&3&  6.9749& 4.164750& 228.386\\
         B&4&  7.1428& 4.621970& 228.406\\
         \bottomrule
    \end{tabular}
    \label{tab:param_table}
\end{table}

\par We investigate parameter stability of the 8 qubits listed in \autoref{tab:param_table} over 95 hours during a single cooldown, where we perform for each qubit a lifetime measurement, a Ramsey experiment, and a state discrimination experiment. \autoref{tab:description} lists the measured parameters, their definitions and the experiments from which they were extracted. A more detailed description of the pulse sequences and parameters extraction can be found in the supplementary information (S.I. I.). For a single qubit, each measurement cycle (lifetime measurement + Ramsey experiment + state discrimination experiment) takes between 64\:s and 130\:s, depending on $T_1$. Qubits on the same chip are probed sequentially. Both chips are characterized in parallel using two separate sets of measurement instruments which are placed outside the dilution refrigerator (DR) in a non-temperature-stabilized environment. Additionally, we monitor the temperature of the mixing chamber (MXC) of the DR using a $\text{RuO}_{2}$-temperature sensor (see \autoref{fig:dr}). The Ramsey experiment was carried out with a fixed detuning $\Delta_f$, introduced by adding a phase to the second pulse, and without modifying the calibrated drive frequency (see S.I. I.B.2 for further details). The results of this 95-hour single-cooldown stability analysis are presented in \autoref{subsec:A}.

\begin{table*}[t]
    \centering
    \caption{Description of performed experiments and extracted parameters}
    \begin{tabular}{lp{5cm}lp{5cm}} 
        \toprule
         Experiment & Description & Parameter & Definition \\
         \midrule
         Lifetime measurement & $\mathrm{X}_{180}$-delay-readout sequence & $T_1$ & Longitudinal relaxation time, time constant of the exponential decay fit.\\
         \midrule
        Ramsey & $\mathrm{X}_{90}$-delay-$\mathrm{X}_{90}$-readout sequence & $T_2^*$ & Effective transverse relaxation time, time constant of the oscillatory decay fit.\\
         &  & $|f_\mathrm{Ramsey}|$ & Absolute value of the frequency of Ramsey oscillations.\\
         \midrule
        state discrimination & \multirow{3}{5cm}{\RaggedRight $N$ times readout of the qubit at ground state, $N$ times readout of the qubit at excited state. When the readout results are plotted in the IQ plane, two blobs are observed (ideally). In our case, $N=2^{12}$.} & $\Delta_\text{m}$ & The distance between the means of the IQ blobs.\\
         &  & $\mathcal{F}$ & Readout fidelity, calculated after state discrimination.\\
         &  & $T_\text{eff}$& Effective temperature of the qubit, calculated based on the results of state discrimination.\\
         \bottomrule
    \end{tabular}
    \label{tab:description}
\end{table*}

\par Beyond evaluating stability over several days and during a single cooldown, we also examine qubit stability over extended timescales spanning multiple cooldowns. We track both resonator and qubit frequencies for two representative qubits, A.1 and B.1, along with their lifetimes $T_1$, across 10 cooldown cycles over more than a year (other qubits were monitored but over fewer cooldowns). Each cooldown lasts between 7 and 30\:days and is followed by a warm-up and a vacuum break of the dilution refrigerator, during which the qubit chips remain inside their cryo-packages in ambient air. The 95-hour measurement corresponds to the fifth cooldown in this series. The results of this long-term stability study are presented in \autoref{subsec:B}.

\section{Results}
\subsection{Single-cooldown stability over ca. 95 h}\label{subsec:A}

\par
\begin{figure*}
    \centering
    \includegraphics[width=0.90\linewidth]{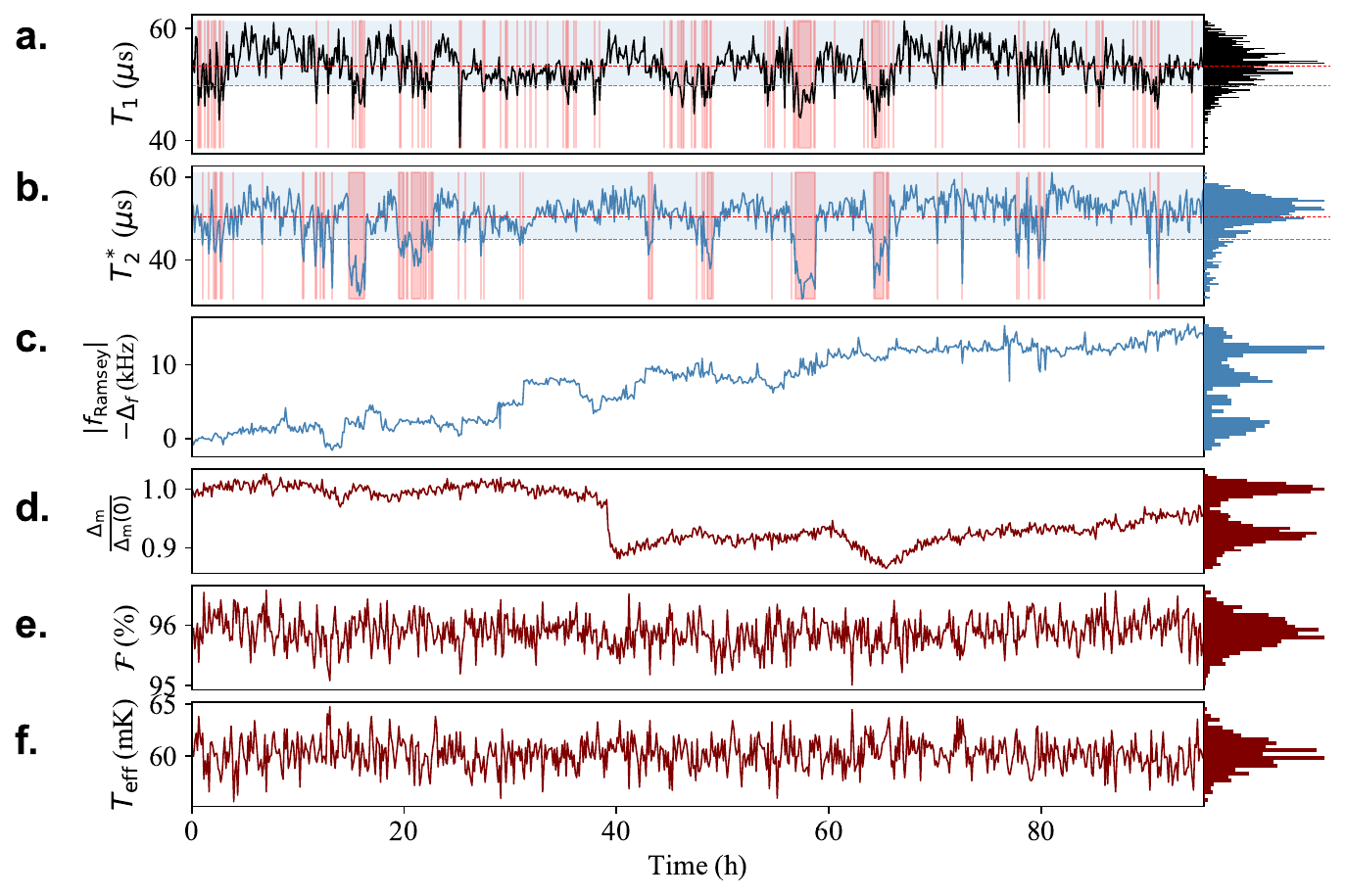}
    \caption{Results of 95-hour single-cooldown measurements for qubit A.2, with time traces on the left and corresponding histograms on the right in each panel: a) $T_1$, b) $T_2^*$, c) $|f_\mathrm{Ramsey}|-\Delta_f$, where $\Delta_f$ is the set detuning, d) $\Delta_\mathrm{m}$, e) readout fidelity $\mathcal{F}$, f) $T_\mathrm{eff}$, and g) $T_\mathrm{MXC}$. The red dotted lines in panels (a) and (b) correspond to the mean values of  $T_1$ and $T_2^*$, respectively. The red-shaded regions indicate where $T_1$ or $T_2^*$ fall by more than one standard deviation below the mean (threshold marked by the dotted horizontal gray lines). These red-shaded regions highlight the simultaneous drops of $T_1$ and $T_2^*$.}
    \label{fig:A.2}
\end{figure*}

 As described in the above section, the stability over 95 hours was analyzed for 8 qubits in total. As an example, \autoref{fig:A.2} shows the results for qubit A.2. The left panels show the time evolution of the measured parameters, and the right panels display their corresponding distribution histograms. The bottom left panel (\autoref{fig:A.2}.g) shows the time evolution of the MXC temperature, $T_{\mathrm{MXC}}$.

\par The histograms of $T_1$ and $T_2^*$ (\autoref{fig:A.2}.a,b) exhibit a negative skewness. This skewness originates from abrupt and aperiodic drops in the time evolution of $T_1$ and $T_2^*$. These drops vary in duration and magnitude and are primarily attributed to spectral diffusion caused by crossings of two-level system (TLS) defects over the qubit frequency \cite{klimov2018fluctuations, burnett2019decoherence, muller2019towards, paik2011observation}. 
The $T_1$ drops coincide most of the time with those of $T_2^*$, as shown by the red intervals in \autoref{fig:A.2}.a,b, and as expected from $1/T_2 = 1/(2T_1)+ 1/T_\mathrm{\varphi}$ and $T_2^*\leq T_2$ \cite{krantz2019quantum,blais2021circuit}.

\par Through $|f_\mathrm{Ramsey}|-\Delta_f$ (\autoref{fig:A.2}.c), we can track the time evolution of the qubit frequency. The qubit frequency exhibits both abrupt jumps as well as drifts. It has been observed in \cite{klimov2018fluctuations} that TLS defects can also undergo either sudden frequency changes (telegraphic) or frequency drifts (diffusive). Since TLS can interact with the qubit and shift its frequency via hybridization \cite{abdurakhimov2022identification,matityahu2024qubit,liu2024observation, chen2025scalable}, the observed qubit frequency changes could be caused by energy changes of nearby TLS \cite{kono2024mechanically}. In total, the observed frequency shift was small (below 15 kHz).

\par
Next, we analyze the state discrimination results. \autoref{fig:A.2}.d shows that $\Delta_\mathrm{m}$ undergoes small but noticeable variations throughout the experiment, which are discussed in the following section. The readout fidelity $\mathcal{F}$ (\autoref{fig:A.2}.e) was relatively stable for A.2 during the measurement period. The effective qubit temperature $T_\text{eff}$ (\autoref{fig:A.2}.f) behaves similarly as the readout fidelity, which is expected because they are related (see S.I. I.B.3).
$T_\text{eff}$ shows no correlation with the steadily decreasing $T_{\mathrm{MXC}}$ (\autoref{fig:A.2}.g).  
This is in agreement with previous findings \cite{scigliuzzo2020primary, lvov2025thermometry, jin2015thermal} that at low mixing-chamber temperatures, the qubit’s effective temperature remains independent of $T_{\mathrm{MXC}}$.

\vspace{1ex}
\noindent\textbf{\small Comparison between qubits}
\vspace{1ex}

\par
\begin{figure*}
    \includegraphics[width=1\linewidth]{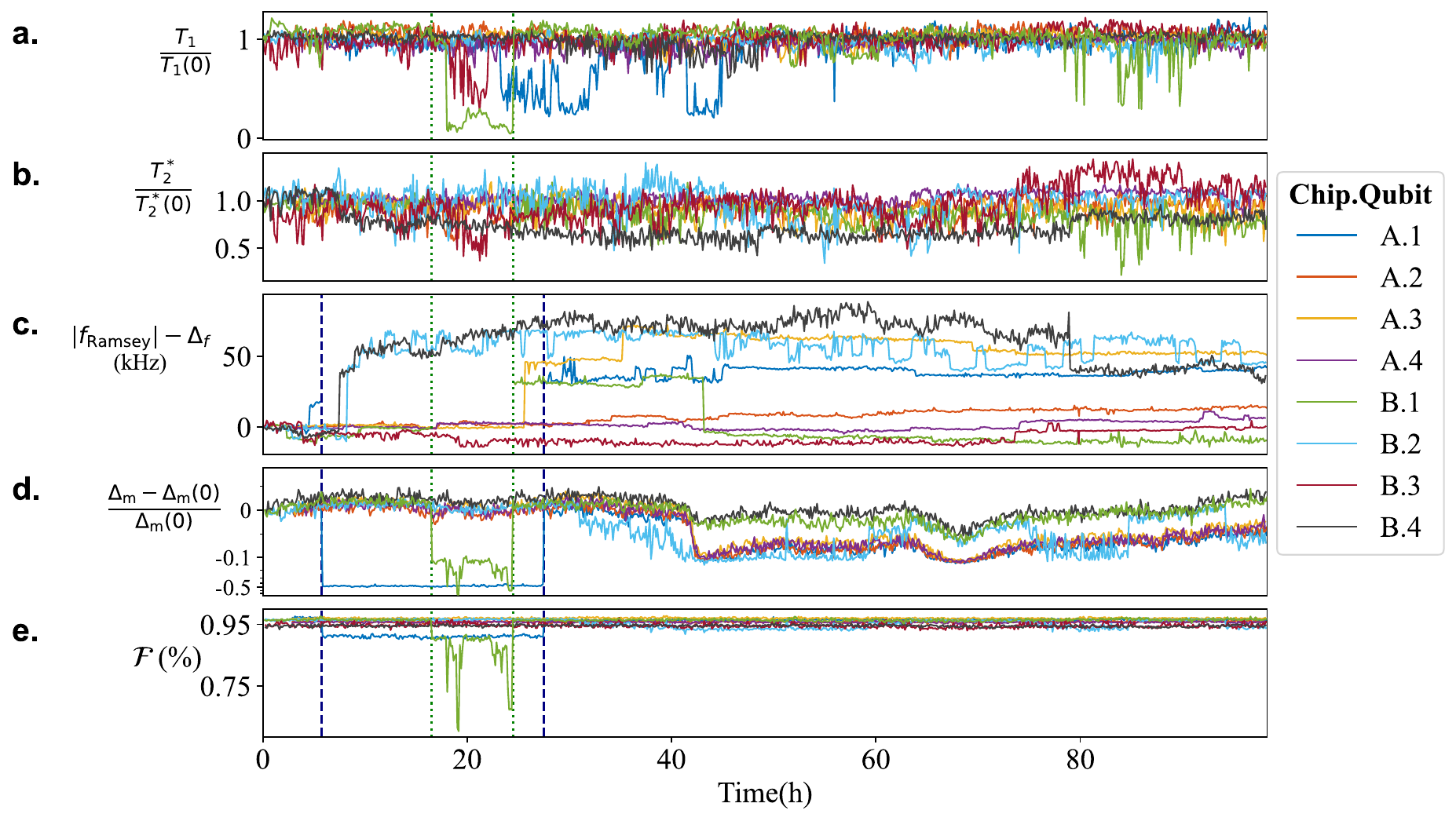}
    \caption{Temporal evolution of a) $T_1(t)$/$T_1(0)$, b) $T_2^*(t)$/$T_2^*(0)$, c) $|f_\mathrm{Ramsey}|-\Delta_f$, d) $\frac{\Delta_\mathrm{m}(t)-\Delta_\mathrm{m}(0)}{\Delta_\mathrm{m}(0)}$, and e) readout fidelity $\mathcal{F}$ for all qubits (except Ramsey results of A.1 and $\Delta_\mathrm{m}$ results for B.3). Panel (d) uses a logarithmic scale on the y-axis with a linear interval $\pm0.1$ around 0. The green dotted lines indicate the time limits of the largest $T_1$ drop observed in qubit B.1, and highlight its impact on the other parameters. Due to $T_2^*$ decrease, the Ramsey results were unreliable during this time interval and therefore omitted. The blue dashed lines delimit the time interval where the readout fidelity and $\Delta_\mathrm{m}$ decreased for A.1 and the Ramsey frequency could not be determined due to $T_2^*$ decrease.}
    \label{fig:correlations}
\end{figure*}

 Next, we investigate how the different parameters compare between different qubits. \autoref{fig:correlations} displays the parameter changes for all measured qubits. 
From $T_2^*$ results, we exclude qubit A.1, due to inadequate detuning during a $T_2^*$ decrease and a too small measurement window after an increase in $T_2^*$ later on. We also omit the $\Delta_\mathrm{m}$ results for B.3, due to strong fluctuations (see S.I. II.C.1).
\par In \autoref{fig:correlations}.a, the $T_1$ time traces show that $T_1$ drops can vary significantly in magnitude across qubits. Some qubits exhibit significant decreases of $T_1$ down to 50\%, or in the worst case scenario down to 10\%, of its original value while others remain stable. As expected, these pronounced $T_1$ drops are mirrored in the $T_2^*$ traces (\autoref{fig:correlations}.b).

\par The most significant decrease affected B.1 (green dotted lines in \autoref{fig:correlations}.a-e). Within this time interval, in addition to $T_2^*$ and $T_1$, $\Delta_\mathrm{m}$ and $\mathcal{F}$ also decreased due to decay during readout.  
Such drastic qubit performance decreases (dropouts) have been previously observed \cite{kono2024mechanically, biznarova2024mitigation, WMI_TUM_17qubit_2024}. They are generally attributed to interactions with strongly coupled TLS and can in principle affect any qubit, occurring stochastically such that extended measurement time spans are more likely to capture these events. Such events can significantly affect QPU operation. Mitigation strategies include decreasing the size of the Josephson junction \cite{colao2025mitigating}, using flux-tunable qubits, or TLS control strategies \cite{you2022stabilizing, chen2025scalable, debroy2024luci, dane2025performance}.

\par We notice that qubit A.1 experienced similar decreases (blue dashed lines in \autoref{fig:correlations}.c-e). However, during the corresponding time interval, no decrease in its $T_1$ was observed. Further investigation is required to explain this behavior.

\par Regarding the qubit frequency (see \autoref{fig:correlations}.c) no correlation with other parameters was observed, except for B.1 and A.1 during specific intervals and B.4, where the main frequency jumps coincide with jumps in $T_2^*$. During the times where we could determine the frequency reliably, $|f_\mathrm{Ramsey}|-\Delta_f$ remains below 100\:kHz, indicating minimal variations of the qubit frequencies and reliable stability.

\par Furthermore, \autoref{fig:correlations}.e demonstrates that the readout fidelity $\mathcal{F}$ remains consistently high for all qubits, apart from A.1 and B.1 during specific intervals, during which also $\Delta_\mathrm{m}$ dropped. Otherwise, changes of $\Delta_\mathrm{m}$ remain small (see \autoref{fig:correlations}.d), reaching a maximum of approximately -13.7\%. \autoref{fig:correlations}.d also reveals that the traces of $(\Delta_\mathrm{m}(t)-\Delta_\mathrm{m}(0))/\Delta_\mathrm{m}(0)$ follow similar temporal profiles across qubits with some offsets. This finding implies that although the baseline values of $\Delta_\mathrm{m}$ differ among qubits, its temporal variations arise from common factors shared between the chips. Since the correlation between the $T_\mathrm{MXC}$ and $\Delta_\mathrm{m}$ is weak and the chips have separate measurement paths, we attribute the $\Delta_\mathrm{m}$ changes to influence(s) of the environment(s) in which components of the measurement setup (wiring, amplifiers, measurement instruments) are placed.

\vspace{1ex}
\noindent\textbf{\small $T_1$ stability}
\vspace{1ex}

\par To benchmark our qubits in terms of $T_1$-stability, we systematically compare results of measurements performed on devices fabricated using similar or different methods. The following section provides, to the best of our knowledge, for the first time a systematic benchmarking of the long-term $T_1$ stability of CMOS-compatible subtractively fabricated qubits compared to lift-off fabricated qubits.

\par For this study, we include only datasets where the original authors provided raw data of $T_1$. We also require that the datasets contain more than 500 points collected over at least 10\:hours and during a single cooldown. These requirements are set to ensure statistical reliability and properly reflect the dynamics of $T_1$. For each qubit, we determine the empirical mean and standard deviation of the corresponding $T_1$ dataset.

\par The considered datasets for the benchmarking use three different fabrication approaches: (i) a CMOS-compatible subtractive process (EMFT, IMEC \cite{van2024advanced}), (ii) the Dolan bridge method (Chalmers \cite{burnett2019decoherence}, Müller et al./IBM \cite{muller2015interacting}), and (iii) the Manhattan method (VTT \cite{tuokkola2025methods}, EPFL \cite{kono2024mechanically}). Both Dolan and Manhattan are lift-off methods. Our own dataset (EMFT) comprises three subsets: EMFT$\_0$, consisting of the 8 qubits analyzed in this work; EMFT$\_1$, with additional chips from the same process run as EMFT$\_0$ \cite{lang2025advancing}; and EMFT$\_2$, with chips from a subsequent run \cite{mayer2025cmos}.
All the mentioned works employ Al/Al-Ox/Al Josephson junctions. All structures other than the Josephson junction are made from aluminum, except for EPFL, and VTT where, e.g., resonators or shunted capacitors are made of Nb.
\par
\begin{figure}
    \centering
    \includegraphics[width=\linewidth]{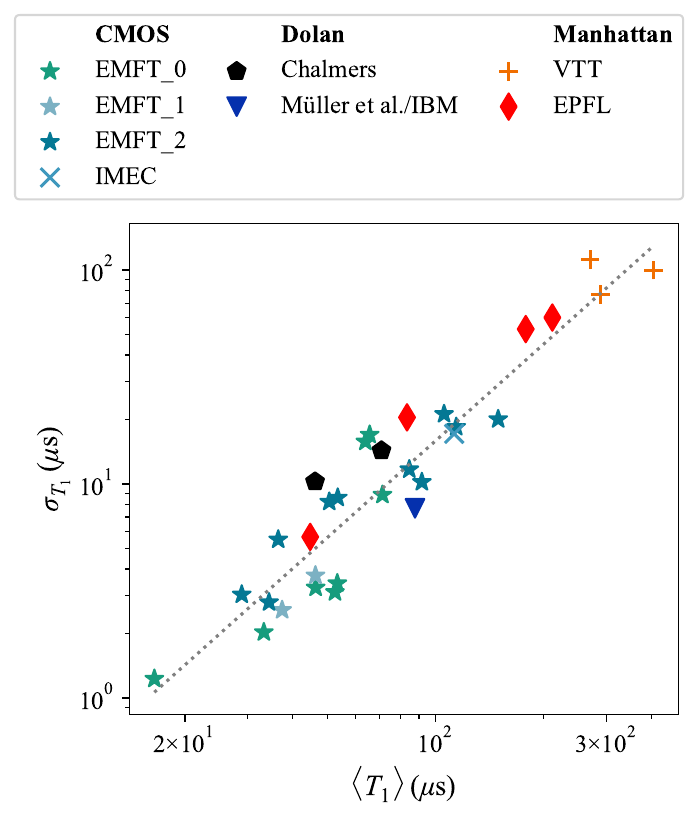}
    \caption{Standard deviation of $T_1$ with respect to its mean value, double-logarithmically scaled. The dotted line represents the fit following \protect\autoref{eq:square_equation} which gives   $a = (15.9\pm0.9)\:\sqrt{\mathrm{Hz}}$.}
    \label{fig:T1 variation} 
\end{figure}

 In \autoref{fig:T1 variation}, we plot the standard deviations $\sigma_{T1}$ with respect to the corresponding average lifetimes $\langle T_1\rangle$. \autoref{fig:T1 variation} shows that all qubits considered follow a common trend where higher average lifetimes $\langle T_1\rangle$ correspond to higher standard deviations $\sigma_{T1}$. This finding is in agreement with previous results \cite{bal2024systematic, burnett2019decoherence, simbierowicz2021qubit} and the theoretical model derived from \cite{you2022stabilizing} where:
\begin{equation}
\begin{aligned}
   \sigma_{T1} \approx a\cdot\langle T_1\rangle^{\frac{3}{2}}
\end{aligned}
\label{eq:square_equation}
\end{equation}
with $a$ being a proportionality factor related to fluctuations of the decay caused by a single TLS (see S.I. II.C.2 for the derivation). This model assumes that TLS defects are the dominant source of energy relaxation \cite{you2022stabilizing}. We find that all qubits, regardless of the fabrication method, can be described with the same proportionality factor $a$. Note that the approximation in \autoref{eq:square_equation} breaks down for $T_1$ values larger than the interval studied here. For details, see S.I. II.C.2. 

\par The observed relation between $\langle T_1\rangle$ and $\sigma_{T1}$, suggests that as the qubit lifetimes increase, the stability decreases, which needs to be taken into account for the QPU operation. In addition, the experimental results show that qubits fabricated with our subtractive CMOS-compatible approach are as stable as superconducting qubits fabricated with the traditional lift-off methods, indicating very similar overall TLS behavior regardless of the fabrication method.

\subsection{Long-term stability (over several cooldowns)}\label{subsec:B}

\par
\begin{figure}
    \centering
    \includegraphics[width=\linewidth]{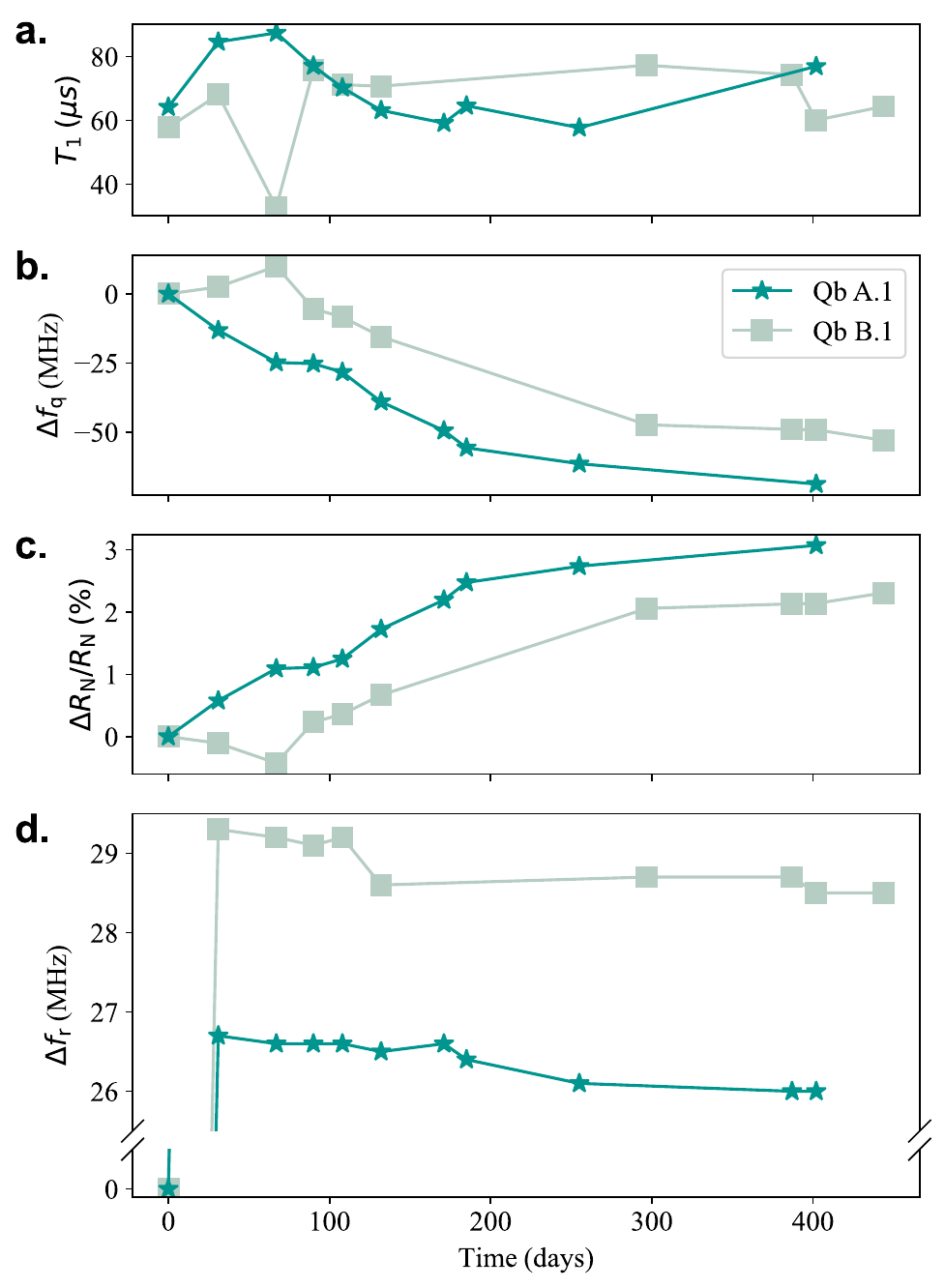}
    \caption{Evolution over multiple cooldown cycles of a) $T_1$ and of b) qubit and d) readout frequencies compared to their initial values ($\Delta f_\text{q}$ and $\Delta f_\text{r}$). c) Evolution of estimated $\Delta R_\mathrm{N}/R_\mathrm{N}$.}
    \label{fig:cooldown_cycles}
\end{figure}

 Two qubits from chips A and B were measured across 10 cooldowns over the course of one year. \autoref{fig:cooldown_cycles} summarizes the results, with each data point corresponding to one cooldown. 

\par Each point in \autoref{fig:cooldown_cycles}.a corresponds to the average $T_1$ for a given cooldown. $T_1$ fluctuates between cooldowns, but remains within bounds around a stable value. This behavior of $T_1$ can be attributed to a reconfiguration of the spectrum of TLS defects due to thermal cycling \cite{de2020two,kim2022effects, colao2025mitigating}.

\par \autoref{fig:cooldown_cycles}.b shows the qubit frequency shifts $\Delta f_\mathrm{q}$ relative to the first cooldown. There is a gradual decrease which appears to saturate toward the end of the measurement series and amounts to around 50-70\:MHz in total. 

The observed decrease in $f_\text{q}$ can be explained by an increase in the room temperature resistance of the junction $R_\text{N}$, as shown by previous aging studies \cite{nesbitt2007time, koppinen2007complete, bilmes2021situ, pop2012fabrication}. 
A higher $R_\text{N}$ reduces the Josephson energy $E_\mathrm{J}\propto 1/R_\text{N}$. Consequently, the transition frequency decreases as follows \cite{koch2007charge}:
\begin{equation}\label{eq:Baratoff}
f_\mathrm{q}
\approx \frac{1}{h}\bigl(\sqrt{8E_\mathrm{J}E_\mathrm{C}}\;-\;E_\mathrm{C}\bigr)
\quad\text{for}\quad
\frac{E_\mathrm{J}}{E_\mathrm{C}}\gg1
\end{equation}
where $E_\mathrm{C}$ is the charging energy of the Josephson junction and $h$ is the Planck constant. The increase in $R_\text{N}$ has been attributed to a gradual, slow, and correlated rearrangement of the structure of the aluminum oxide barrier at room temperature. Over time, traps states at the metal-oxide interface relax and reconfigure towards deeper potential minima, thereby increasing the barrier thickness \cite{nesbitt2007time}. If the barrier is contaminated with resist residues or hydrates, the aging and variation of the resistance and qubit frequency are more pronounced \cite{pop2012fabrication, koppinen2007complete, bilmes2021situ}.

\par To quantify the variation that could have occurred to $R_\mathrm{N}$, we calculate it based on our values of $f_\mathrm{q}$ and the following expression derived from \autoref{eq:Baratoff}, and $E_\mathrm{J}=h\Delta/(8e^2R_\mathrm{N})$ \cite{koch2007charge, ambegaokar1963tunneling}: 
\begin{equation}
    R_\mathrm{N}=\frac{h\Delta E_\mathrm{C}}{e^2(f_\mathrm{q}\cdot h+E_\mathrm{C})^2}.
\end{equation}
$\Delta$ is the superconducting gap and $e$ is the elementary charge.
Using our previously determined critical temperature $T_\mathrm{c}$ \cite{lang2025advancing} to calculate $\Delta$, and the values of anharmonicities in \autoref{tab:param_table}, we obtain the results shown in \autoref{fig:cooldown_cycles}.c. 

On average, the measured qubits maintained $\Delta R_\mathrm{N}/R_\mathrm{N}$ below 3.4\% over 400 days, which is lower than the aging observed for junctions fabricated with lift-off methods \cite{van2024advanced, nesbitt2007time, kim2022effects, koppinen2007complete}. We acknowledge that direct comparison between different technologies and setups is challenging due to variations in the time elapsed between fabrication completion and initial characterization (approximately 27 days in our case), as well as the slowing-down of aging during cooldown \cite{nesbitt2007time}. We hypothesize that the low aging we observe from our qubits is due to the resist removal step using $O_2$-plasma at 250\,°C \cite{lang2025advancing, koppinen2007complete}.

\par Similarly to $\Delta f_\mathrm{q}$, \autoref{fig:cooldown_cycles}.d shows the resonator frequency shift $\Delta f_\mathrm{r}$ with respect to the first cooldown. There is a pronounced jump between the first and second cooldowns for both qubits. All qubits on chips A and B experience this jump in their readout frequencies (see S.I. III). We also confirm that this change affected the test resonators (see \autoref{fig:layout}) and the bare readout resonance frequencies. Interestingly, this jump was limited to chips A and B; other chips from the same wafer show only slight frequency decreases. The origin of this initial shift remains unclear. Subsequent decreases of $f_\mathrm{r}$ in later cooldowns can be caused by changes of the bare resonator frequency $f_\mathrm{r,bare}$ as well as changes of $f_\mathrm{q}$ and the qubit-resonator coupling $g$ \cite{naghiloo2019introduction,greene2023calibration, gao2021practical,d2024characterization}:
\begin{equation}
    f_\mathrm{r} \approx f_\mathrm{r,bare}+ \frac{g^2}{f_\mathrm{r,bare}-f_q}.
\end{equation} 
As $g\propto(1/R_\mathrm{N})^{1/4}$ and $f_q\propto(1/R_\mathrm{N})^{1/2}$, an increase of $R_\mathrm{N}$ leads to a decrease of $f_\mathrm{r}$. We estimate that this effect accounts for roughly 25\,\% of the observed changes, and 75\,\% from decreases of $f_\mathrm{r,bare}$. The latter might stem from an increase of the effective dielectric permittivity $\varepsilon_{\mathrm{eff}}$ possibly due to material changes, or an increase in $l_\mathrm{tot}$ due to cracks formation and/or growth in the aluminum resonator induced by thermal cycling \cite{syed2017effect}. These changes affect $f_\mathrm{r,bare}$ following the equation \cite{luo2025demonstration,li2023experimentally}:
\begin{equation}
    f_\mathrm{r,bare}  \approx\frac{c_0}{4 l_\mathrm{tot}\cdot\sqrt{\varepsilon_{\mathrm{eff}}}}
\end{equation}
$c_0$ denotes the vacuum light speed.

\section{Summary and discussion}
\par Our results from measurements over a 95-hour period and during a single cooldown demonstrate that subtractively fabricated CMOS-compatible superconducting qubits can achieve parameter stability on par with lift-off devices, while offering high yield, long lifetimes, and the scalability advantages of industry-grade CMOS processes \cite{lang2025advancing,mayer2025cmos}. The observed scaling $\sigma_{T1}\propto \langle T_{1}\rangle^{3/2}$ combined with our observations on $T_1$ temporal evolution confirms that the TLS-qubit interactions remain the limiting factor for $T_1$ stability, even at elevated lifetimes. We believe that the difference in $T_1$ stability between the studied fabrication methods is minimal (\autoref{fig:T1 variation}) because oxide formation at the relevant interfaces is nearly identical across them, while the main difference is the patterning (lift-off uses EBL, CMOS-compatible subtractive method uses optical lithography). The additional vacuum break needed in the subtractive approach also does not exacerbate the TLS problem, according to our findings. The variations in the fabrication methods do not, therefore, seem to translate into the TLS ensemble governing $T_1$-instabilities.
Furthermore, in relation to TLS-caused fluctuations, we acknowledge the possibility of $T_1$-dropouts in superconducting qubits, which can heighten the error probability in a QPU. Beyond $T_1$, $T_2^*$, and $T_2$, TLS defects can also shift the qubit frequency or indirectly degrade other performance metrics (e.g., lead to increased decay during readout, reducing readout fidelity). These findings underscore the need for improved materials and surface treatments that mitigate the TLS effects, or for stabilizing strategies that circumvent existing TLS \cite{you2022stabilizing, chen2025scalable, debroy2024luci, dane2025performance}. Additional mechanisms may also contribute to instabilities of some qubit parameters.
\par Our measurements over 10 cooldowns spanning more than a year show that $T_1$ (in mean value) is subject to fluctuations due to thermal cycling but retains a stable baseline. Qubit frequencies $f_\mathrm{q}$ experience a gradual decrease, driven by an increase in the room temperature resistance $R_{N}$ of the Josephson junction. 
However, this aging effect is moderate: the computed $R_{N}$ values show an average $\Delta R_{N}/R_{N}$ of only 3.4\% over 400 days. Nevertheless, such variations should ideally be minimized or accounted for in future QPUs undergoing multiple thermal cycles.\\\\
\par In summary, our benchmarking and long-term stability studies validate the viability of subtractively fabricated CMOS-compatible superconducting qubits for scalable quantum processors, while pinpointing the key challenges that must be addressed to achieve large-scale fault-tolerant devices.\\

\begin{acknowledgments}
\par The authors would like to thank Tianmu Zhang and Lukas Sigl form Zurich Instruments for their helpful support in setting up the qubit measurement system. We acknowledge helpful discussions and assistance with test chip design and setting up of cryogenic measurements from G. Huber, I. Tsitsilin, F. Haslbeck and C. Schneider from the Quantum Computing group at the Walther Meissner Institute. We also appreciate the whole Fraunhofer EMFT clean room staff for the professional fabrication.
\par This work was funded by the Munich Quantum Valley (MQV) – Consortium Scalable Hardware and Systems Engineering (SHARE), funded by the Bavarian State Government with funds from the Hightech Agenda Bavaria, the Munich Quantum Valley Quantum Computer Demonstrator - Superconducting Qubits (MUNIQC-SC) 13N16188, funded by the Federal Ministry of Education and Research, Germany, and the Open Superconducting Quantum Computers (OpenSuperQPlus) Project - European Quantum Technology Flagship.
\end{acknowledgments}

\balancecolsandclearpage

\setcounter{equation}{0}
\setcounter{figure}{0}
\setcounter{table}{0}
\setcounter{page}{1}
\setcounter{section}{0}
\setcounter{NAT@ctr}{0}
\makeatletter
\renewcommand{\fnum@figure}{Fig.~\thefigure}
\renewcommand{\fnum@table}{Tab.~\thetable}

\renewcommand{\theequation}{S\arabic{equation}}
\renewcommand{\thefigure}{S\arabic{figure}}
\renewcommand{\thetable}{S\arabic{table}}
\renewcommand{\bibnumfmt}[1]{[S#1]} 
\renewcommand{\citenumfont}[1]{S#1}
\makeatother
\onecolumngrid
\begin{center}
\textbf{\large Supplementary information: Stability studies on subtractively-fabricated CMOS-compatible superconducting transmon qubits}
\vspace{3em}
\end{center}
\twocolumngrid

\section{Experiments}
\par In this section, we refer to Refs. \cite{Skrantz2019quantum,zhinst2023dynamicpulses,naghiloo2019introductionmeasurestuff}
\subsection{Pulses}
\par To drive the qubits, we use pulses with Gaussian envelope. First, the pulse length is set (140\:ns-200\:ns), then the amplitude is varied in a Rabi experiment to calibrate the $X_{180}$ and $X_{90}$ gates. Readout uses rectangular pulses. We optimize the readout by performing state discrimination experiment for different readout powers and choosing the parameters that maximize readout fidelity.
\subsection{Pulse sequences and parameters extraction}
\par \autoref{fig:pulses} illustrates the pulse sequences for lifetime measurement, Ramsey experiment, and state discrimination measurement. Each readout is followed by a passive reset (not shown in \autoref{fig:pulses}). In the following, we describe how the different parameters in Tab. II of the main text are extracted. 
\subsubsection{Lifetime measurement}
\par For each delay $\tau$, the pulse sequence shown in \autoref{fig:pulses}.a,b is executed on the qubit. After averaging over $2^{10}$ repetitions, we obtain $P_{|1\rangle}(\tau)$, the excited-state population after delay $\tau$ following excitation. By fitting $P_{|1\rangle}(\tau)$ to an exponential decay:
\begin{equation}
    P_{|1\rangle}(\tau)=A\cdot \mathrm{exp}\left(-\frac{\tau}{T_1}\right)+B
\end{equation}
we obtain the lifetime of the qubit $T_1=1/\Gamma_1$. $A$ and $B$ are fitting parameters and $\Gamma_1$ is the energy decay rate. Typically, $A\approx1$ and $B\approx0$.
\subsubsection{Ramsey experiment}
\par The Ramsey pulse sequence displayed in \autoref{fig:pulses}.c,d yields an oscillatory decay of $P_{|1\rangle}(\tau)$ that follows the equation:
\begin{equation}
    P_{|1\rangle}(\tau)=A\cdot\mathrm{cos}\left(2\pi f_\mathrm{Ramsey}+\phi_0\right)\mathrm{exp}\left(-\frac{\tau}{T_2^*}\right)+B
\end{equation}
where $A$, $B$, and $\phi_0$ are fitting parameters, $T_2^*$ is the effective transverse relaxation time, and $f_\mathrm{Ramsey}$ is the Ramsey oscillation frequency. Typically, $A\approx B\approx 1/2$ and $\phi_0\approx0$.
\par Using $f_\mathrm{Ramsey}=f_\mathrm{q}-f_\mathrm{Drive}$, where $f_\mathrm{q}$ and $f_\mathrm{Drive}$ are the qubit and the drive frequencies, respectively, we can calibrate the drive to the qubit frequency with high accuracy. However, due to the symmetry of the cosine function and $\phi_0\approx0$, a single Ramsey experiment ($X_{90}$-delay $\tau$-$X_{90}$-readout) cannot determine the sign of $f_\mathrm{Ramsey}$. This problem can be solved by either using a pulse sequence where the second pulse is $Y_{90}$, i.e. $X_{90}$-delay $\tau$-$Y_{90}$-readout, or by introducing an additional detuning $\Delta_f$. Assuming the initial drive frequency is $f_\mathrm{Drive,i}=f_\mathrm{q}+\delta f$ and $\Delta_f>|\delta f|$, the possible values $f_\mathrm{Ramsey}$ takes are $f_\mathrm{Ramsey}=|\delta f|+\Delta_f$ or $f_\mathrm{Ramsey}=-|\delta f|+\Delta_f$, which are both positive. The calibrated drive frequency $f_\mathrm{Drive, cal}$ is then:
\begin{equation}
    f_\mathrm{Drive, cal}=f_\mathrm{Drive,i}-\left(f_\mathrm{Ramsey}-\Delta_f\right)\approx f_q
\end{equation}
The introduced detuning also benefits our experiment, where we track the time evolution of $f_\mathrm{Ramsey}$: Since the time span of a single experiment $t_\mathrm{max}$ is limited by decay or choice,$f_\mathrm{Ramsey}$  smaller than $1/t_\mathrm{max}$ cannot be detected. With detuning, frequency shifts are modulated around $\Delta_f$, increasing sensitivity to small drifts relative to the initial frequency. The detuning $\Delta_f$ is implemented by adding a delay-dependent phase $\phi(\tau)$ to the second pulse of the Ramsey sequence \cite{ramsey1950molecular,shuker2019ramseyphi,roos2008quantum_ramsey_ion}: 
\begin{figure*}
\centering
\begin{minipage}{0.45\textwidth}\RaggedRight{\textbf{a.}}\\
\centering
\begin{tikzpicture}[scale=1.0]
\draw[->] (0,0) -- (6,0) node[right]{t};

\path[fill=pulsecolor,opacity=0.3] plot[domain=0:1.2,smooth,variable=\x] 
  ({\x}, {1.6*exp(-((\x-0.6)^2)/0.1)}) -- (1.2,0) -- (0,0) -- cycle;
\draw[domain=0:1.2,smooth,variable=\x,pulsecolor,thick] 
plot ({\x}, {1.6*exp(-((\x-0.6)^2)/0.1)});
\node[above] at (0.6,1.7) {$X_{180}$};

\path[fill=gray,opacity=0.3] plot[gray,thick] (3.0,0) rectangle (4.0,0.5);
\draw[gray,thick] (3.0,0) rectangle (4.0,0.5);
\node[above] at (3.5,0.6) {Readout};

\draw (0,0) -- (0,-0.1) node[below] {$0$};
\draw (1.2,0) -- (1.2,-0.1) node[below] {$t_{\text{gate}}$};
\draw (3.0,0) -- (3.0,-0.1) node[below] {$\tau+t_{\text{gate}}$};
\end{tikzpicture}
\end{minipage}
\hfill
\begin{minipage}{0.45\textwidth}\RaggedRight{\textbf{b.}}\\
\centering

\[
\Qcircuit @C=1em @R=1em {
\lstick{|0\rangle} & \gate{X} & \qw & \gate{\mathrm{Delay}(\tau)} & \meter \\
}
\]
\end{minipage}

\vspace{1em} 
\begin{minipage}{0.45\textwidth}\RaggedRight{\textbf{c.}}\\
\centering
\begin{tikzpicture}[scale=1.0]
\draw[->] (0,0) -- (6,0) node[right]{t};

\path[fill=pulsecolor,opacity=0.3] plot[domain=0:1.2,smooth,variable=\x] 
  ({\x}, {0.8*exp(-((\x-0.6)^2)/0.1)}) -- (1.2,0) -- (0,0) -- cycle;
\draw[domain=0:1.2,smooth,variable=\x,pulsecolor,thick] 
plot ({\x}, {0.8*exp(-((\x-0.6)^2)/0.1)});
\node[above] at (0.6,0.9) {$X_{90}$};

\path[fill=pulsecolor,opacity=0.3] plot[domain=3.0:4.2,smooth,variable=\x] 
  ({\x}, {0.8*exp(-((\x-3.6)^2)/0.1)}) -- (1.2,0) -- (0,0) -- cycle;
\draw[domain=3.0:4.2,smooth,variable=\x,pulsecolor,thick] plot ({\x}, {0.8*exp(-((\x-3.6)^2)/0.1)});
\node[above] at (3.6,0.9) {$X_{90}$};

\path[fill=gray,opacity=0.3] plot[gray,thick] (4.3,0) rectangle (5.3,0.5);
\draw[gray,thick] (4.3,0) rectangle (5.3,0.5);
\node[above] at (4.8,0.6) {Readout};

\draw (0,0) -- (0,-0.1) node[below] {$0$};
\draw (1.2,0) -- (1.2,-0.1) node[below] {$t_{\text{gate}}$};
\draw (3.0,0) -- (3.0,-0.1) node[below] {$\tau+t_{\text{gate}}$};
\end{tikzpicture}
\end{minipage}
\hfill
\begin{minipage}{0.45\textwidth}\RaggedRight{\textbf{d.}}\\
\centering

\[
\Qcircuit @C=1em @R=1em {
\lstick{|0\rangle} & \gate{X_{90}} & \qw & \gate{\mathrm{Delay}(\tau)} & \gate{R_z(\phi)} & \qw & \gate{X_{90}} & \qw & \meter\\
}
\]
\end{minipage}
\vspace{1em} 

\begin{minipage}{0.45\textwidth}
\centering
\RaggedRight{\textbf{e.}}
\[
\left(
\begin{tikzpicture}[scale=1.0, baseline =(current bounding box.center) ]
\draw[->] (0,0) -- (1.2,0) node[right]{t};
\path[fill=gray,opacity=0.3] plot[gray,thick] (0,0) rectangle (1,0.5);
\draw[gray,thick] (0,0) rectangle (1,0.5);
\node[above] at (0.5,0.6) {Readout};
\draw (0,0) -- (0,-0.1) node[below] {$0$};
\end{tikzpicture}\right)\times N
\quad
\left(
\begin{tikzpicture}[scale=1.0, baseline =(current bounding box.center)]
\draw[->] (0,0) -- (2.5,0) node[right]{t};
\path[fill=pulsecolor,opacity=0.3] plot[domain=0:1.2,smooth,variable=\x] 
  ({\x}, {1.6*exp(-((\x-0.6)^2)/0.1)}) -- (1.2,0) -- (0,0) -- cycle;
\draw[domain=0:1.2,smooth,variable=\x,pulsecolor,thick] 
plot ({\x}, {1.6*exp(-((\x-0.6)^2)/0.1)});
\node[above] at (0.6,1.7) {$X_{180}$};
\path[fill=gray,opacity=0.3] plot[gray,thick] (1.25,0) rectangle (2.25,0.5);
\draw[gray,thick] (1.25,0) rectangle (2.25,0.5);
\node[above] at (1.75,0.6) {Readout};
\draw (0,0) -- (0,-0.1) node[below] {$0$};
\end{tikzpicture}\right)\times N\]
\end{minipage}
\hfill
\begin{minipage}{0.45\textwidth}\RaggedRight{\textbf{f.}}\\
\centering

\[
\left.\begin{array}{c}
\Qcircuit @C=1em @R=0.5em {
& \lstick{\ket{0}} & \qw & \meter\\
& \lstick{\ket{0}} & \qw & \meter\\
& \lstick{\vdots} & & \vdots\\\\
& \lstick{\ket{0}} & \qw & \meter\\
}
\end{array}
\right\} \times N 
\quad
\left.\begin{array}{c}
\Qcircuit @C=1em @R=0.5em {
& \lstick{\ket{0}} & \qw &\gate{X}& \meter \\
& \lstick{\ket{0}} & \qw &\gate{X}& \meter\\
& \lstick{\vdots} & & \vdots & \vdots \\\\
& \lstick{\ket{0}} & \qw &\gate{X}& \meter\\
}
\end{array}
\right\} \times N
\]

\end{minipage}

\caption{Pulse sequences for qubit characterization. Left panels show pulse-level implementation; right panels show equivalent quantum circuit representations. a,b) lifetime measurement. c,d) Ramsey sequence with virtual $R_z(\phi)$ for frequency detuning. e,f) Single-shot readout with N repetitions of $|0\rangle$ and $|1\rangle$ measurements}
\label{fig:pulses}
\end{figure*}
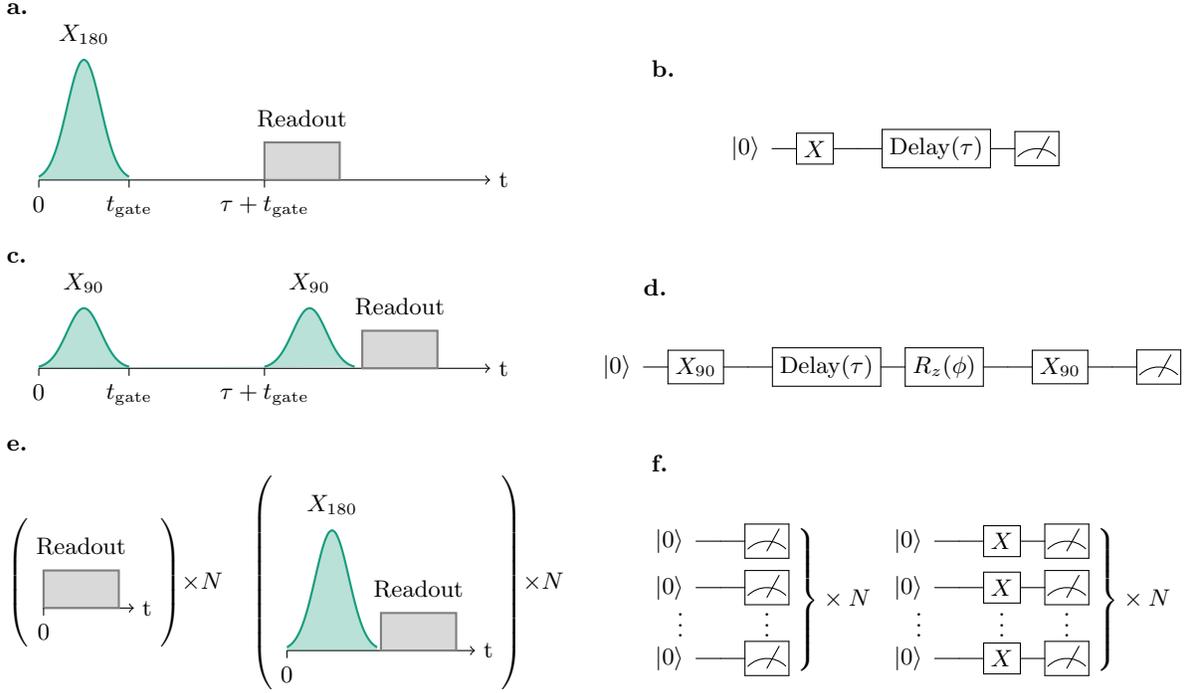
\begin{equation}
    \phi(\tau)=2\pi\Delta_f\cdot\tau~\mathrm{mod}(2\pi)
\end{equation}
This operation is denoted in \autoref{fig:pulses}.d by the gate $R_z(\phi)$.
According to the Nyquist theorem, undersampling is possible if $f_q$ shifts significantly \cite{Shannon}. However, the time span must be large enough to capture $T_2^*$ accurately, creating a trade-off between measurement speed and sampling rate. Ideally, the sampling rate and detuning would dynamically adapt to the evolution of $f_\mathrm{Ramsey}$ (not implemented in this work). \autoref{tab:sampling} presents the measurements parameters for each qubit. The time spans are chosen to satisfy $t_\mathrm{max}\approx5T_2^*$.
\begin{table}[]
    \centering
    
    \begin{tabular}{c c c c c}
         \toprule
        Qubit&$t_\mathrm{max}$&$f_\mathrm{Nyquist}$&Detuning&$\langle T_2^*\rangle$\\
        \midrule
        A.1&120&166.7&20&28.56\\
        A.2&250&80&10&51.26\\
        A.3&200&100&10&38.88\\
        A.4&150&133.3&10&34.96\\
        \midrule
        B.1&150&133.3&15&39.38\\
        B.2&250&80&10&44.15\\
        B.3&250&80&15&44.03\\
        B.4&100&200&15&15.96\\
        \bottomrule     
    \end{tabular}
    \caption{Ramsey measurement parameters. Time values are given in $\mu s$ and frequency values are given in kHz. For A.1, only the first valid values of $ T_2^*$ were averaged. For the other qubits, $\langle T_2^*\rangle$ is obtained by fitting the distributions. The dropout of B.1 was excluded from the fit.}   
    \label{tab:sampling}
\end{table}
\subsubsection{State discrimination experiment}
Using the setup depicted in \autoref{fig:pulses}.e,f, we obtain two blobs in the IQ plane, e.g. \autoref{fig:A3_IQ}.a, each corresponding to a qubit state. We compute the centers (means) of the two blobs and the distance separating them, $\Delta_\mathrm{m}$. To calculate the readout fidelity $\mathcal{F}$ and effective qubit temperature $T_\mathrm{eff}$, we use in this work Quadratic Discriminant Analysis (QDA), whereby the IQ plane is divided into two regions, each representing a state. The readout fidelity $\mathcal{F}$ is then \cite{magesan2015machine,Schen2023transmon}:\\
\begin{equation}
\begin{aligned}
\mathcal{F}&=\frac{P\left(|1\rangle\big||1\rangle\right)+P\left(|0\rangle\big||0\rangle\right)}{2}\\
&=1-\frac{P\left(|0\rangle\big||1\rangle\right)+P\left(|1\rangle\big||0\rangle\right)}{2}\\
\end{aligned}
\end{equation}
where $P\left(|i\rangle\big||j\rangle\right)$ is the probability of finding the qubit in the region corresponding to the state $|i\rangle$ after being prepared to the state $|j\rangle$.
The effective qubit temperature $T_\mathrm{eff}$ can be estimated from $P\left(|1\rangle\big||0\rangle\right)$ and $P\left(|0\rangle\big||0\rangle\right)$. After preparing the qubit in the state $|0\rangle$, the transition $|0\rangle\rightarrow|1\rangle$ is possible, e.g. due to thermal excitation by photons at the qubit frequency $f_\mathrm{q}$ or hot non-equilibrium quasiparticles. We can determine the effective qubit temperature through the Boltzmann-Maxwell distribution: \cite{valenzuela2007microwave,serniak2018hot, Slvov2025thermometry,Sjin2015thermal}:
\begin{equation}
    \frac{P\left(|1\rangle\big||0\rangle\right)}{P\left(|0\rangle\big||0\rangle\right)}=\mathrm{exp}\left(-\frac{hf_\mathrm{q}}{k_\mathrm{B}T_\mathrm{eff}}\right)
\end{equation}
which gives:
\begin{equation}
    T_\mathrm{eff}=-\frac{hf_\mathrm{q}}{\mathrm{k}_\mathrm{B}\cdot \mathrm{ln}\left(\frac{P\left(|1\rangle\big||0\rangle\right)}{P\left(|0\rangle\big||0\rangle\right)}\right)}
    \label{eq:BM}
\end{equation}
where $h$ is the Planck constant and $\mathrm{k}_\mathrm{B}$ is the Boltzmann constant.
The term $P\left(|0\rangle\big||0\rangle\right)$ in \autoref{eq:BM} can be approximated as $P\left(|0\rangle\big||0\rangle\right)=1-P\left(|1\rangle\big||0\rangle\right)\approx 1$ since typically $P\left(|1\rangle\big||0\rangle\right)\ll1$. An increase in $P\left(|1\rangle\big||0\rangle\right)$  reflects in a decrease of $\mathcal{F}$ and an increase of $T_\mathrm{eff}$, giving an anti-correlation between $T_\mathrm{eff}$ and $\mathcal{F}$.
\section{Single-cooldown stability}
    
\subsection{Results from qubits other than A.2}
\par In this section, we present additional results in a manner similar to Fig.3 of the main manuscript. \autoref{fig:A.1} corresponds to qubit A.1, where the Ramsey results were omitted due to the aforementioned strong abrupt frequency change. \autoref{fig:A.3} and \autoref{fig:A.4} display the measurement results for qubits A.3 and A.4, respectively. \autoref{fig:B.1} shows the temporal evolution of parameters of B.1, with the time interval of the strongest dropout indicated by dotted lines. \autoref{fig:B.2} shows the measurements results for B.2, with the particularity of a correlation between $\Delta_\mathrm{m}$ and $\mathcal{F}$. In \autoref{fig:B.3} corresponding to B.3, we can observe the fluctuations in $\Delta_\mathrm{m}$ (see \autoref{subsubsec:transmon}). \autoref{fig:B.4} displays the evolution of B.4 parameters. Interestingly, the two major frequency jumps coincide with jumps in $T_2^*$.
\subsection{Sources of fluctuations}
\subsubsection{Measurement-induced state transitions}\label{subsubsec:transmon}
\begin{figure}
    \centering
    \begin{minipage}{\linewidth}\RaggedRight{\textbf{a.}}
            \includegraphics[width=\linewidth]{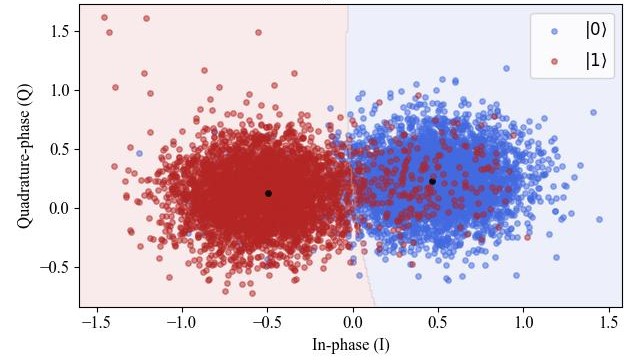}
            \label{fig:normal}
            
    \end{minipage}\hfill
    \begin{minipage}{\linewidth}\RaggedRight{\textbf{b.}}
            \includegraphics[width=\linewidth]{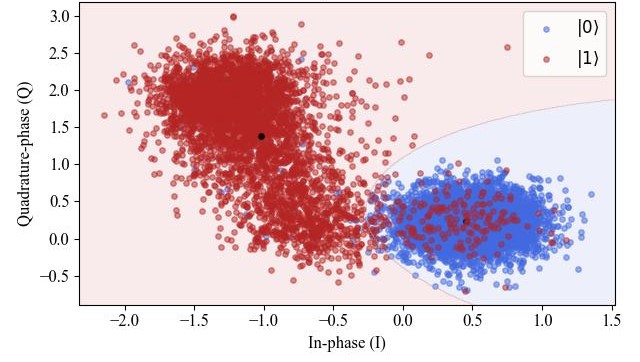}
            \label{fig:banana}
            
    \end{minipage}
    \caption{Two state discrimination measurements of qubit B.3. a) Measurement with two well-defined blobs (time (N) in \autoref{fig:B.3}) corresponding to states $|0\rangle$ (blue) and $|1\rangle$ (red), representing the expected result for a two-level-system. The black dots represent the centers of the blobs used to calculate $\Delta_\mathrm{m}$. b) Measurement-induced state transitions (time (I) in \autoref{fig:B.3}), where the red points (excited state) spread over the IQ plane, instead of forming a Gaussian blob. After state discrimination, we find that the result for $\Delta_\mathrm{m}$ is almost double compared with (a).}
    \label{fig:A3_IQ}
\end{figure}
\par In this section, we discuss the fluctuations in $\Delta_\mathrm{m}$ experienced by B.3, which could be observed in \autoref{fig:B.3}. \autoref{fig:A3_IQ}.a and \autoref{fig:A3_IQ}.b show the results of two separate state discrimination measurements performed on B.3: the first with a relatively low $\Delta_\mathrm{m}$ taken at the time (N), and the second with high $\Delta_\mathrm{m}$ taken at the time (I). Times (I) and (N) are indicated in \autoref{fig:B.3}. In this comparison, \autoref{fig:A3_IQ}.a with low $\Delta_\mathrm{m}$ corresponds to the expected result for a two-level system. In \autoref{fig:A3_IQ}.b, we observe that measurement points prepared in the excited state forming almost 3 connected blobs in the IQ plane, moving the blobs centers away from each other and increasing $\Delta_\mathrm{m}$. This behavior could be attributed to measurement-induced state transitions\cite{connolly2025full, dumas2024measurement, Sshillito2022dynamics}, also referred to as ionization\cite{dumas2024measurement, Sshillito2022dynamics}, which occurs when certain numbers of photons are present in the resonator cavity. In this scenario, the spread of the excited state blob could be the result of either the qubit decaying back to the first excited state during readout or hybridization (superposition of the ionized state and the first excited state). We hypothesize that the sudden switching between the two cases (\autoref{fig:A3_IQ}.a and \autoref{fig:A3_IQ}.b) and consequently the $\Delta_\mathrm{m}$ fluctuations could be due to fluctuations in the photon number in the readout cavity.
\subsubsection{TLS-qubit interaction}
\par The model introduced in Ref.\cite{Syou2022stabilizing} states the following equations:
\begin{equation}
    \langle\Gamma_{1,\mathrm{tot}}\rangle = N_\mathrm{TLS}\langle \Gamma_{1,\mathrm{single\_TLS}}\rangle
    \label{eq1}
\end{equation}
\begin{equation}
    \mathrm{Var}(\Gamma_{1,\mathrm{tot}}) =N_\mathrm{TLS}\mathrm{Var}(\Gamma_{1,\mathrm{single\_TLS}})
    \label{eq2}
\end{equation}
where $N_\mathrm{TLS}$ is the TLS density, $\Gamma_{1,\mathrm{single\_TLS}}$ is the decay rate of the qubit under the influence of a single TLS, and $\Gamma_{1,\mathrm{tot}}$ is the decay rate due to a TLS ensemble. Var denotes the variance. To derive the approximation in Eq. 1 of the main text, we use the Taylor expansion of distribution $T_{1,\mathrm{tot}}=1/\Gamma_{1,\mathrm{tot}}$ around the mean $\langle\Gamma_{1,\mathrm{tot}}\rangle$ to the second-term:
\begin{equation}
\begin{aligned}
    T_{1,\mathrm{tot}}&= 1/ \Gamma_{1,\mathrm{tot}}\\
    &\approx \frac{1}{\langle\Gamma_{1,\mathrm{tot}}\rangle}-\frac{\Gamma_{1,\mathrm{tot}}-\langle\Gamma_{1,\mathrm{tot}}\rangle}{\langle\Gamma_{1,\mathrm{tot}}\rangle^2}\\&+\frac{(\Gamma_{1,\mathrm{tot}}-\langle\Gamma_{1,\mathrm{tot}}\rangle)^2}{\langle\Gamma_{1,\mathrm{tot}}\rangle^3}
\end{aligned}
\label{eq:taylor}
\end{equation}
We also use the Delta method:
\begin{equation}
    \mathrm{Var}(T_{1,\mathrm{tot}})\approx\mathrm{Var}(\Gamma_{1,\mathrm{tot}})/\langle\Gamma_{1,\mathrm{tot}}\rangle^4
    \label{eq:4}
\end{equation}
\autoref{eq:taylor} yields:
\begin{equation}
\begin{aligned}
\langle T_{1,\mathrm{tot}}\rangle&\approx \frac{1}{\langle\Gamma_{1,\mathrm{tot}}\rangle}+\frac{\mathrm{Var}(\Gamma_{1,\mathrm{tot}})}{\langle\Gamma_{1,\mathrm{tot}}\rangle^3}
\label{eq:mean}
\end{aligned}
\end{equation}
From \autoref{eq1} and \autoref{eq2}, we can write:
\begin{equation}
\begin{aligned}
     \mathrm{Var}(\Gamma_{1,\mathrm{tot}}) &\approx\frac{\mathrm{Var}(\Gamma_{1,\mathrm{single\_TLS}})}{\langle \Gamma_{1,\mathrm{single\_TLS}}\rangle}\langle\Gamma_{1,\mathrm{tot}}\rangle\\&:=\beta\langle\Gamma_{1,\mathrm{tot}}\rangle
\end{aligned}
\end{equation}
Substituting in \autoref{eq:mean}, we obtain:
\begin{equation}
\begin{aligned}
     \langle T_{1,\mathrm{tot}}\rangle&\approx \frac{1}{\langle\Gamma_{1,\mathrm{tot}}\rangle}+\frac{\beta}{\langle\Gamma_{1,\mathrm{tot}}\rangle^2}
     \label{eq:approx}
\end{aligned}
\end{equation}
Solving this second-degree equation for $\langle\Gamma_{1,\mathrm{tot}}\rangle$ and eliminating the negative solution gives:
\begin{equation}
\begin{aligned}
     \langle\Gamma_{1,\mathrm{tot}}\rangle\approx\frac{1+\sqrt{1+4\beta \langle T_{1,\mathrm{tot}}\rangle}}{2\langle T_{1,\mathrm{tot}}\rangle}
\end{aligned}
\end{equation}
Consequently:
\begin{equation}
\begin{aligned}
    \mathrm{Var}(T_{1,\mathrm{tot}})&\approx\frac{\beta}{\langle\Gamma_{1,\mathrm{tot}}\rangle^3}\\
    &\approx\beta\left(\frac{2\langle T_{1,\mathrm{tot}}\rangle}{1+\sqrt{1+4\beta \langle T_{1,\mathrm{tot}}\rangle}}\right)^3\\
\end{aligned}
\end{equation}
which gives:
\begin{equation}
\begin{aligned}
    \sigma_{T1}&=\sqrt{\mathrm{Var}(T_{1,\mathrm{tot}})}\\&\approx\sqrt{\beta}\left(\frac{2\langle T_{1,\mathrm{tot}}\rangle}{1+\sqrt{1+4\beta \langle T_{1,\mathrm{tot}}\rangle}}\right)^\frac{3}{2}
\end{aligned}
\label{complex_model}
\end{equation}

\begin{figure}
    \centering
    \includegraphics[width=0.95\linewidth]{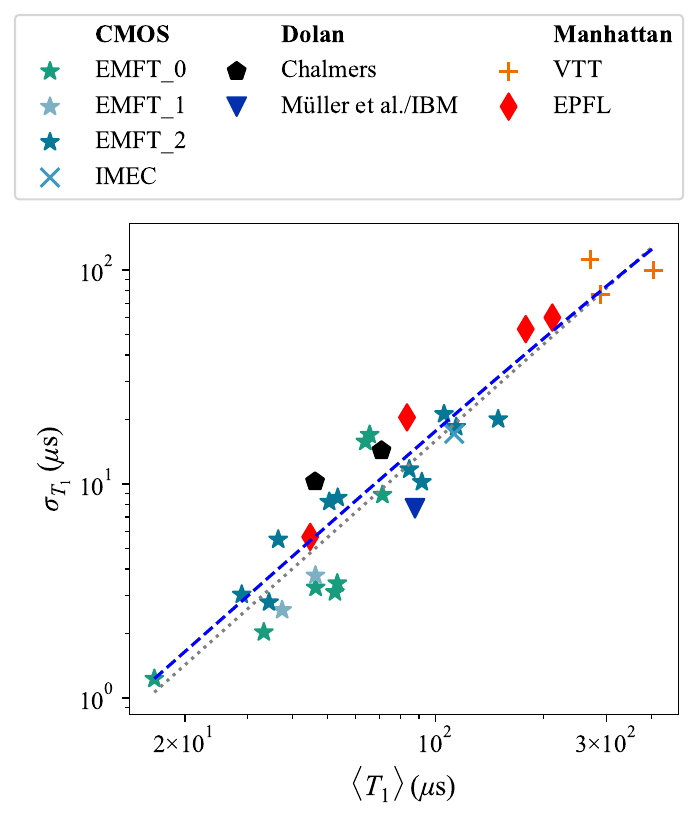}
    \caption{Standard deviation of $T_1$ with respect to its mean value, double-logarithmically scaled. The dotted gray line represents the fit following \protect\autoref{eq:square_equation} which gives $a = (15.9\pm0.9)\:\sqrt{\mathrm{Hz}}$. The dashed blue line is the fit following \protect\autoref{complex_model} which gives $\beta = (343\pm47)\:\mathrm{Hz}$.}
    \label{fig:t1_benchmark}
\end{figure}
By fitting the data to \autoref{complex_model} as shown in \autoref{fig:t1_benchmark} (blue-dashed), we find $\beta=343\pm47\,$Hz. For $\beta \ll \langle\Gamma_{1,\mathrm{tot}}\rangle$, e.g. $\langle\Gamma_{1,\mathrm{tot}}\rangle\geq 3400$\,Hz$\approx 1/294\,\mu$s, \autoref{eq:approx} could be re-written to:
\begin{equation}
\begin{aligned}
     \langle T_{1,\mathrm{tot}}\rangle&\approx \frac{1}{\langle\Gamma_{1,\mathrm{tot}}\rangle}
\end{aligned}
\end{equation}
which gives the approximation used in the main manuscript:
\begin{equation}
    \sigma_{T1}\approx\sqrt{\beta}\langle T_{1,\mathrm{tot}}\rangle^\frac{3}{2}:=a\langle T_{1,\mathrm{tot}}\rangle^\frac{3}{2}
    \label{eq:square_equation}
\end{equation}
The fitted values for $a$ and $\sqrt{\beta}$ differ by $\sim16.5\%$, which is expected since the fit for $a$ includes some points at the threshold of the approximation.  By increasing $\langle T_{1,\mathrm{tot}}\rangle$ further (e.g. beyond 1\,ms), the superlinear regime ($\sigma_{T1}\propto\langle T_{1,\mathrm{tot}}\rangle^\frac{3}{2}$) no longer holds and in addition, a higher-order Taylor expansion would be required (compare \autoref{eq:taylor}). However, $\sigma_{T1}$ still increases for higher $\langle T_{1,\mathrm{tot}}\rangle$. A possible way to improve $T_1$-stability is by influencing $\beta$, which Ref. \cite{dane2025performance} proved possible.
\section{Long-term stability (over several cooldowns)}
\begin{figure}
    \centering
    \includegraphics[width=\linewidth]{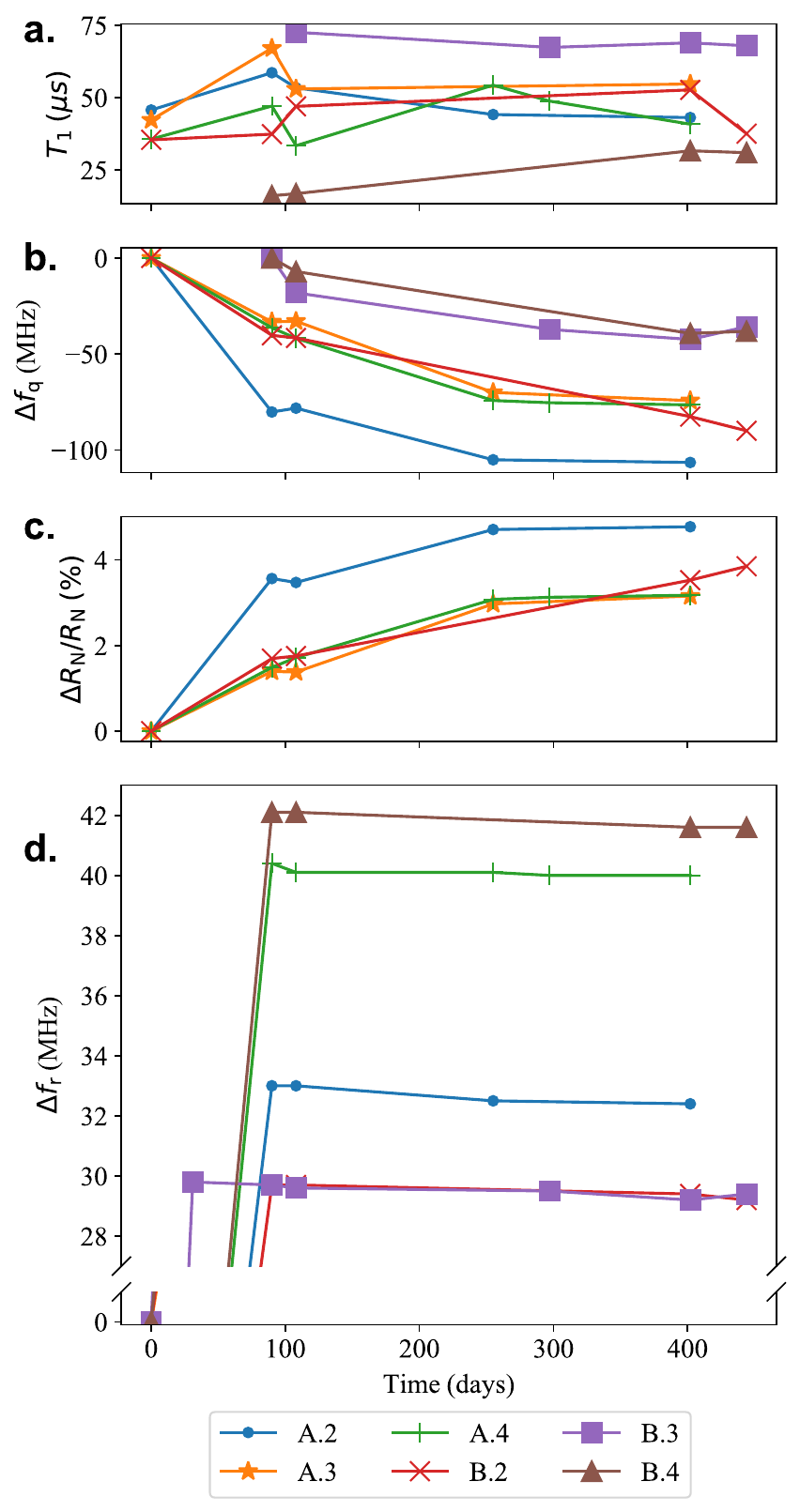}
    \caption{Evolution over multiple cooldown cycles of a) $T_1$ and of b) qubit frequency shifts $\Delta f_\text{q}$ and d) readout frequency shifts $\Delta f_\text{r}$ for qubits A.2-4 and B.2-4. c) Evolution of estimated $\Delta R_\mathrm{N}/R_\mathrm{N}$. Frequency and $T_1$measurements for qubits B.3 and B.4 during the first cooldown are unavailable. That is why both qubits are not handled in panel (c). The zero-time point refers to the same first cooldown in Fig.6 of the main text.}
    \label{fig:aging}
\end{figure}
\par In \autoref{fig:aging} we present the rest of the aging data for qubits other than A.1 and B.1. B.3 and B.4 frequency and $T_1$ measurements during the first cooldown are not available. The remaining qubits exhibit similar behavior as described in the main text.

\balancecolsandclearpage
\begin{figure*}
    \centering
    \includegraphics[width=0.9\linewidth]{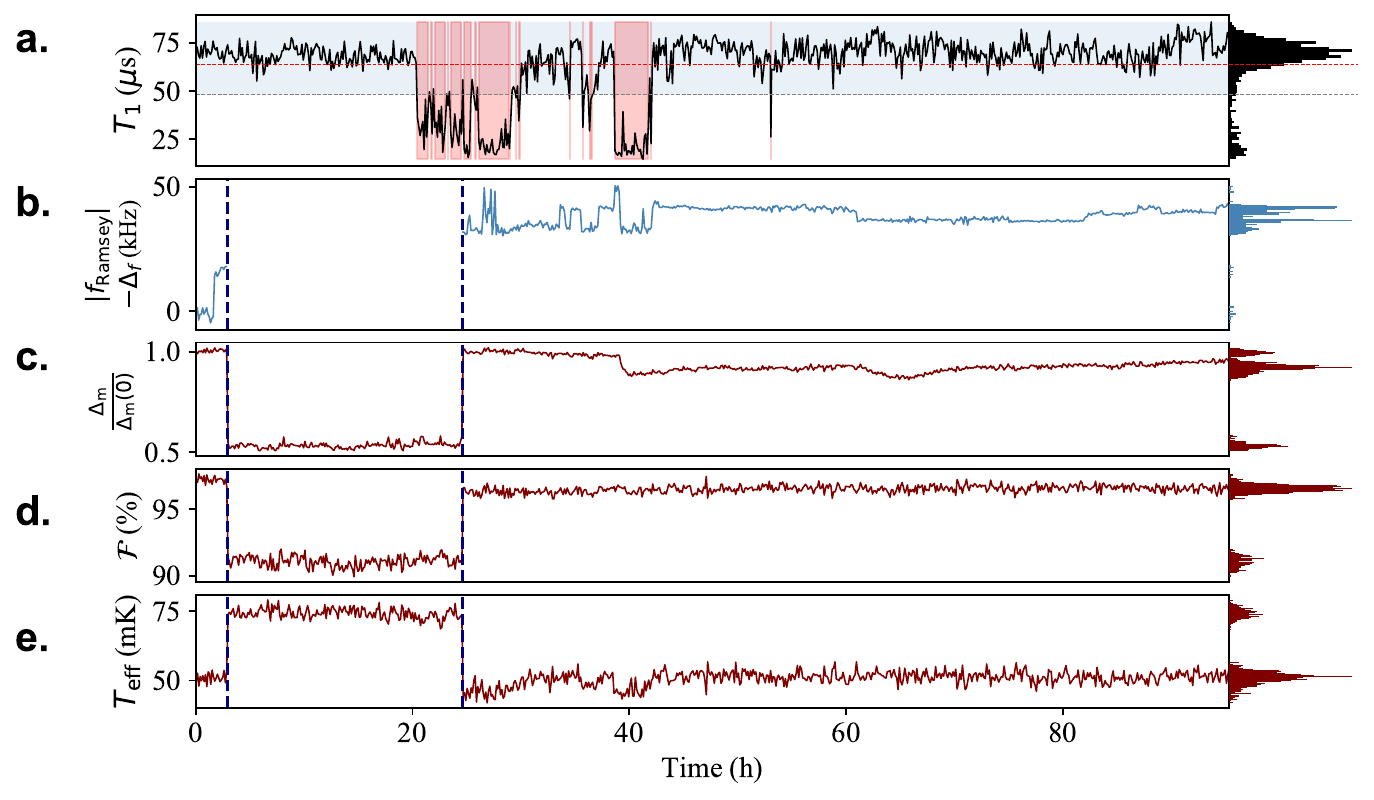}
    \caption{Temporal evolution of qubit A.1 parameters. $T_2^*$ measurements are omitted due to inadequate detuning or measurement time span, making the fits of $T_2^*$ unreliable. The blue dashed lines highlight the period where $T_2^*$ dropped significantly, in addition to deterioration of the readout and the effective temperature. Due to the $T_2^*$ decrease, the frequency could not be fitted during this period.}
    \label{fig:A.1}
\end{figure*}
\begin{figure*}
    \centering
    \includegraphics[width=0.9\linewidth]{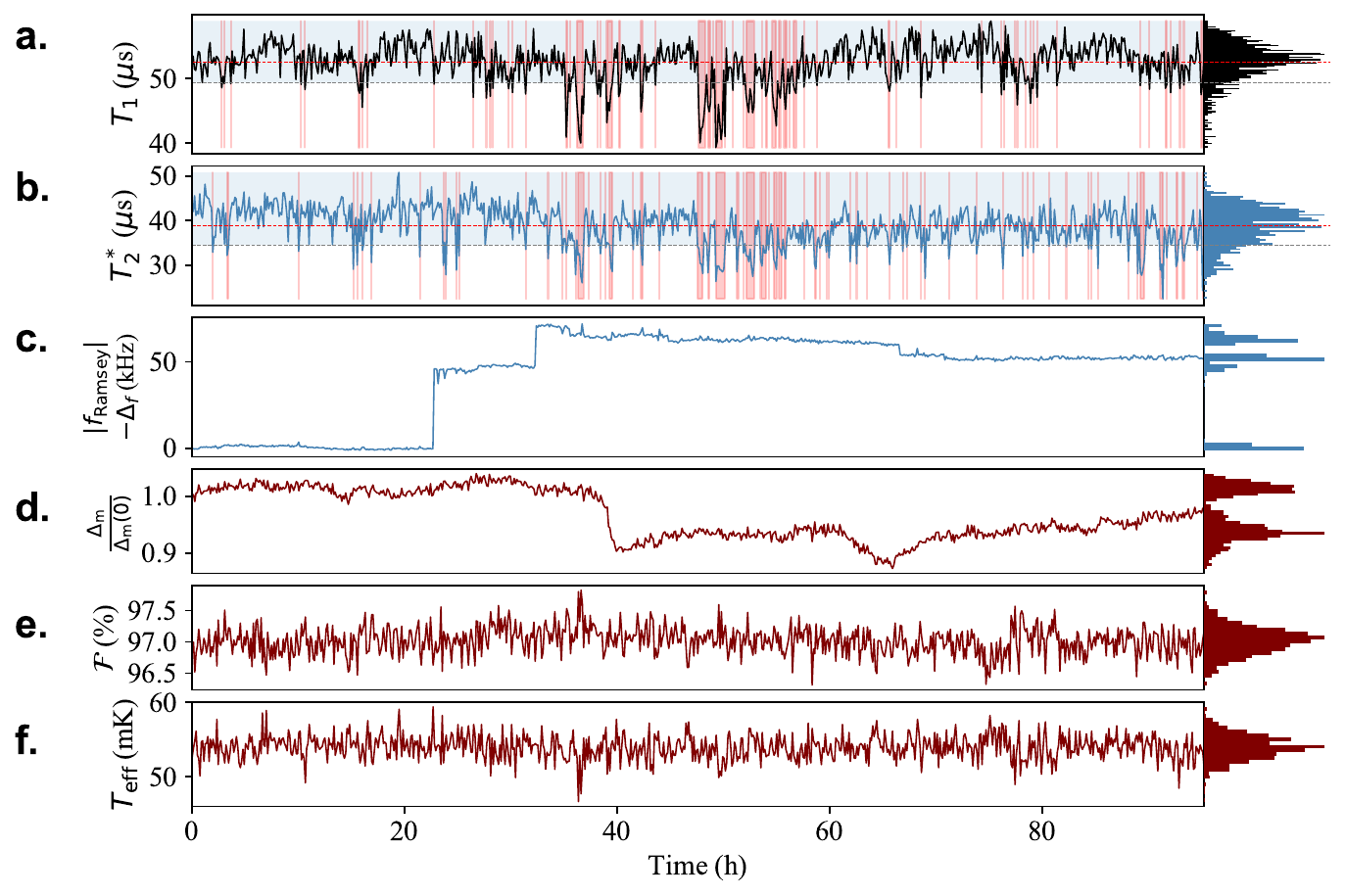}
    \caption{Temporal evolution of qubit A.3 parameters}
    \label{fig:A.3}
\end{figure*}
\begin{figure*}
    \centering
    \includegraphics[width=0.9\linewidth]{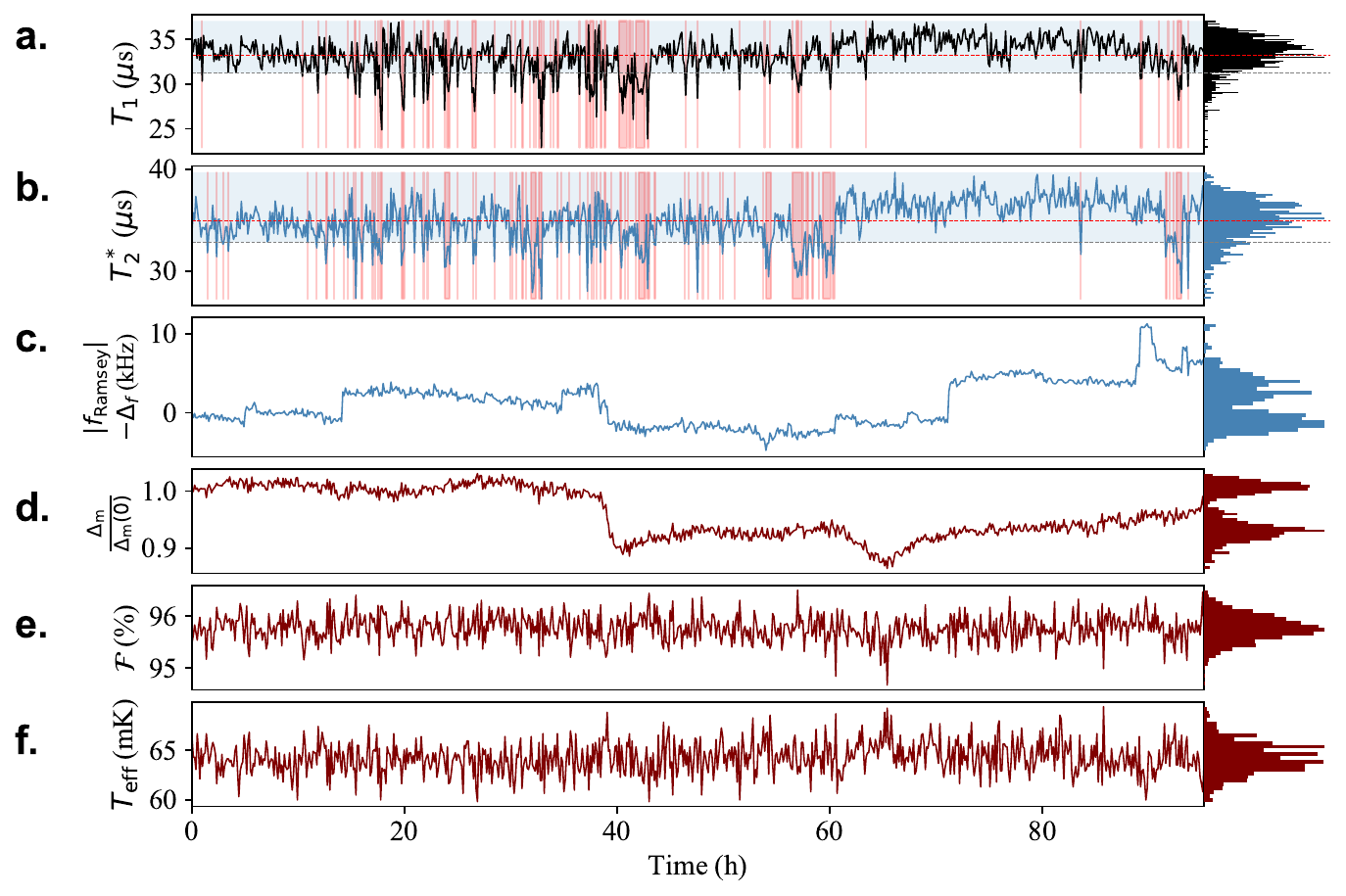}
    \caption{Temporal evolution of qubit A.4 parameters}
    \label{fig:A.4}
\end{figure*}
\begin{figure*}
    \centering
    \includegraphics[width=0.9\linewidth]{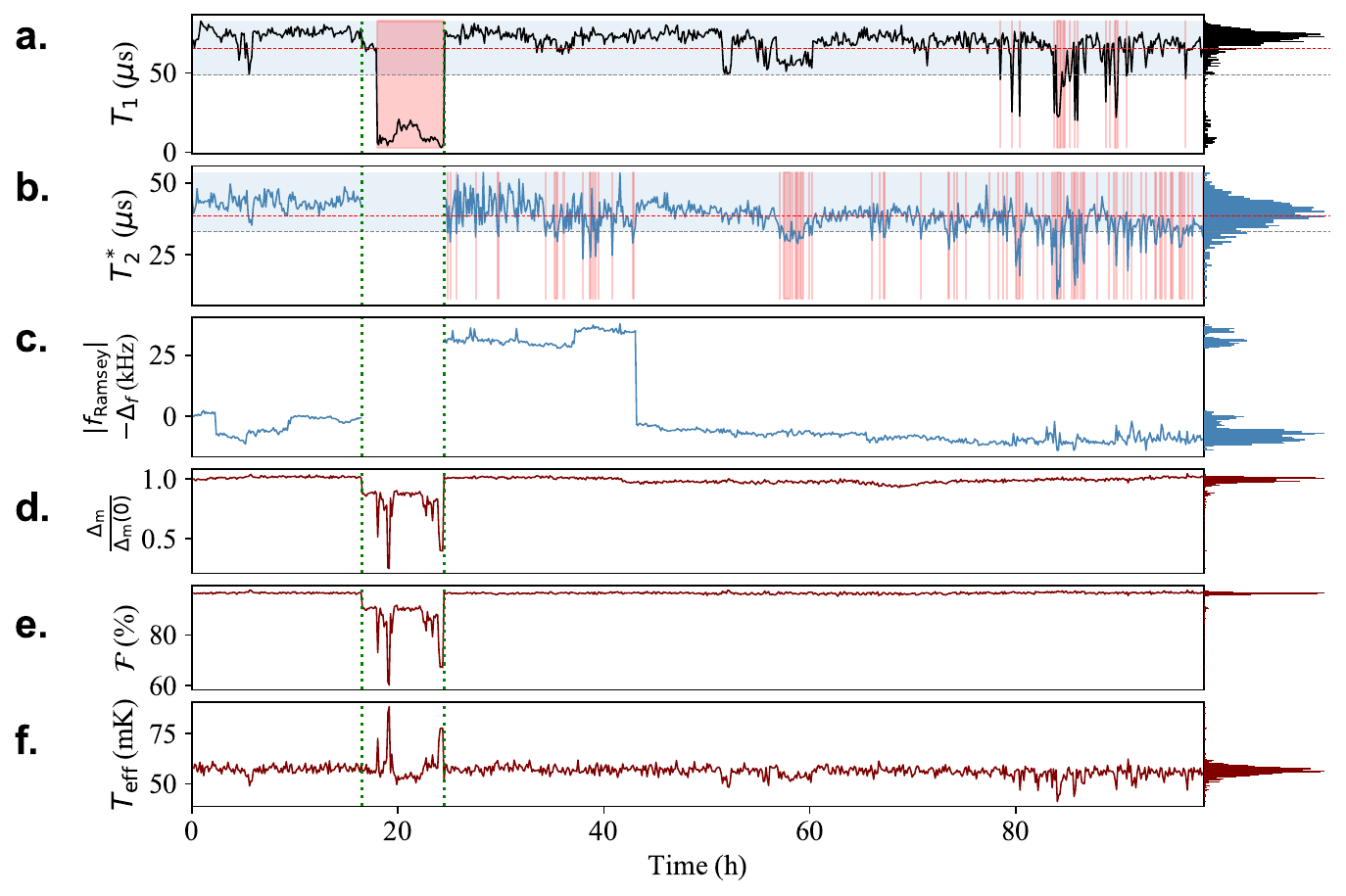}
    \caption{ Temporal evolution of qubit B.1 parameters. Dotted lines indicate the time interval of the strongest dropout.}
    \label{fig:B.1}
\end{figure*}
\begin{figure*}
    \centering
    \includegraphics[width=0.9\linewidth]{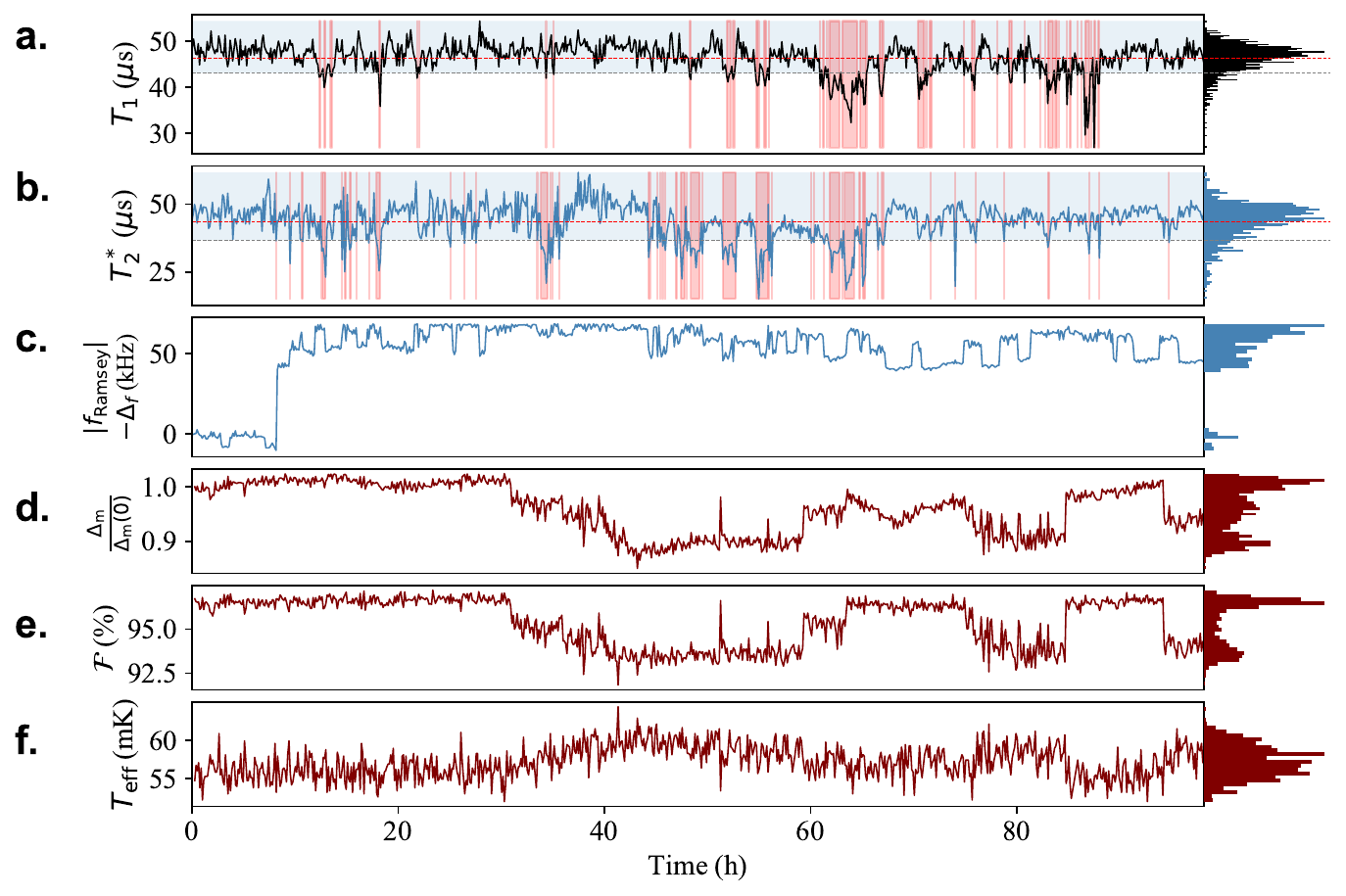}
    \caption{Temporal evolution of qubit B.2 parameters. Note the correlation between $\Delta_\mathrm{m}$ and $\mathcal{F}$.}
    \label{fig:B.2}
\end{figure*}
\begin{figure*}
    \centering
    \includegraphics[width=0.9\linewidth]{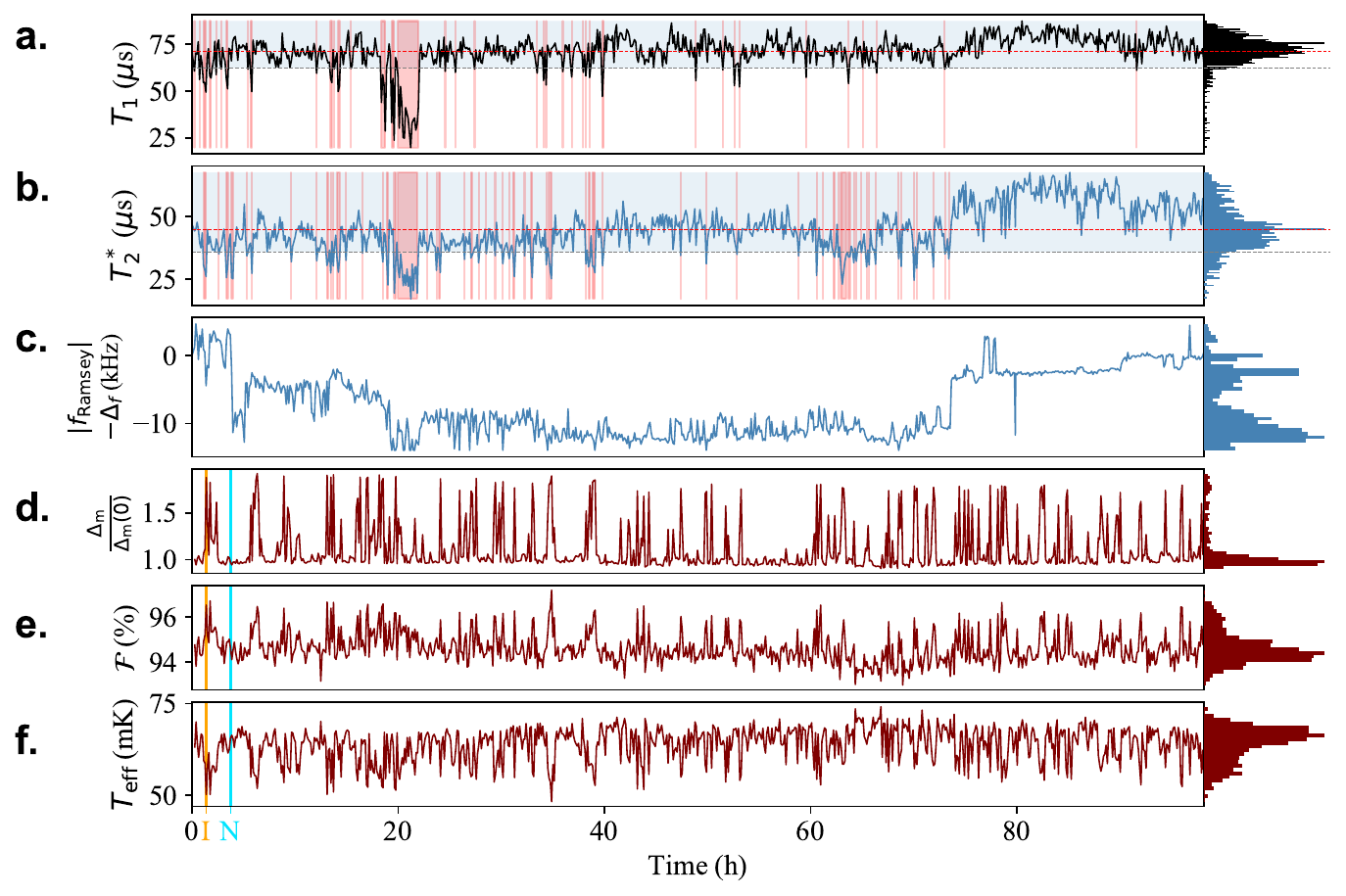}
    \caption{Temporal evolution of qubit B.3 parameters, with strong fluctuations in $\Delta_\mathrm{m}$, with the time (N) and (I) marked in light-blue and orange and corresponding to \protect\autoref{fig:A3_IQ}.a and \autoref{fig:A3_IQ}.b respectively.}
    \label{fig:B.3}
\end{figure*}
\begin{figure*}
    \centering
    \includegraphics[width=0.9\linewidth]{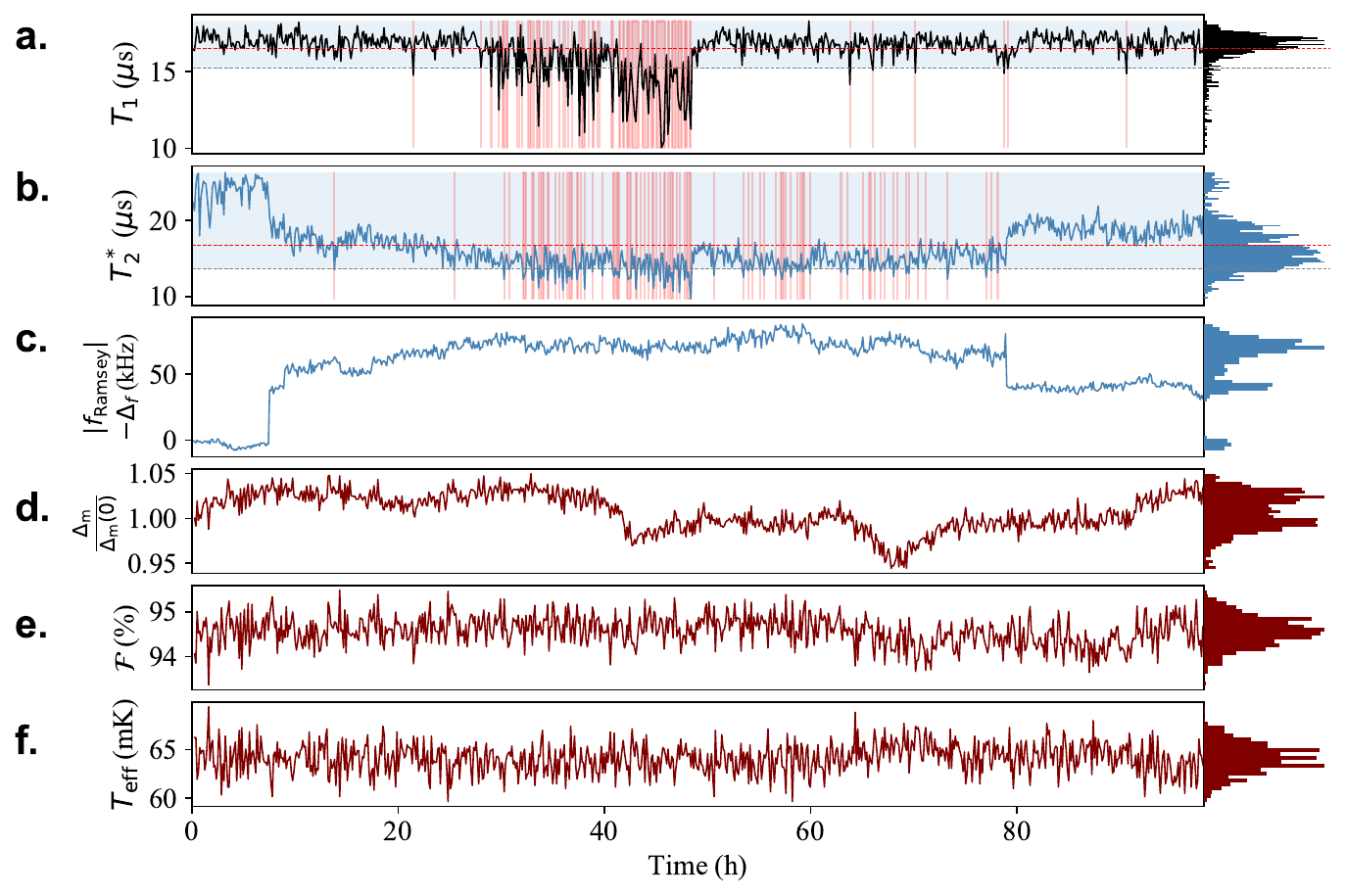}
    \caption{Temporal evolution of qubit B.4 parameters. The two major frequency jumps coincide with jumps in $T_2^*$}
    \label{fig:B.4}
\end{figure*}

\end{document}